\documentclass[superscriptaddress,groupedaddress,nofootnoteinbib,11pt]{article}
\pdfoutput=1 
\usepackage{jheppub}

\usepackage[utf8x]{inputenc}
\usepackage{mathtools,slashed,mathrsfs}
\usepackage[caption=false]{subfig}
\usepackage{dcolumn}
\usepackage{multirow}
\usepackage{tabularx}
\usepackage{booktabs}
\usepackage{bm}
\usepackage{comment}
\usepackage{setspace}
\usepackage[dvipsnames]{xcolor}
\usepackage[normalem]{ulem} 
\usepackage{enumerate}
\usepackage{siunitx}

\newcommand{\Sec}[1]{Sec.~\ref{#1}}

\newcommand{\App}[1]{Appendix~\ref{#1}}
\newcommand{\Fig}[1]{Fig.~\ref{#1}}
\newcommand{\Eq}[1]{Eq.~(\ref{#1})}
\newcommand{\Eqs}[2]{Eqs.~(\ref{#1}) and (\ref{#2})}
\newcommand{\Eqst}[2]{Eqs.~(\ref{#1})--(\ref{#2})}

\newcommand{\beq}{\begin{equation}}
\newcommand{\eeq}{\end{equation}}
\newcommand{\ba}{\begin{array}}
\newcommand{\ea}{\end{array}}
\newcommand{\bea}{\begin{eqnarray}}
\newcommand{\eea}{\end{eqnarray} }
\newcommand{\be}{\begin{eqnarray}}
\newcommand{\ee}{\end{eqnarray}}
\newcommand{\bal}{\begin{align}}
\newcommand{\eal}{\end{align}}
\newcommand{\bi}{\begin{itemize}}
\newcommand{\ei}{\end{itemize}}
\newcommand{\ben}{\begin{enumerate}}
\newcommand{\een}{\end{enumerate}}
\newcommand{\bc}{\begin{center}}
\newcommand{\ec}{\end{center}}
\newcommand{\bt}{\begin{table}}
\newcommand{\et}{\end{table}}
\newcommand{\btb}{\begin{tabular}}
\newcommand{\etb}{\end{tabular}}

\newcommand{\bl}{\left}
\newcommand{\br}{\right}
\newcommand{\ie}{\textit{i.e.}\ }

\newcommand\nn{\nonumber}

\newcommand{\eV}{\mathrm{eV}}
\newcommand{\keV}{\mathrm{keV}}

\newcommand{\GeV}{\mathrm{GeV}}

\newcommand{\Mpc}{\mathrm{Mpc}}

\newcommand{\HOunit}{\mathrm{km/s/Mpc}}

\newcommand{\dd}{\ensuremath{\mathrm{d}}}

\newcommand{\Lag}{\mathcal{L}}
\newcommand{\mH}{\mathcal H}

\newcommand{\eq}{\ensuremath{\mathrm{eq}}}
\newcommand{\dec}{\ensuremath{\mathrm{dec}}}

\newcommand{\lcdm}{\ensuremath{\Lambda\mathrm{CDM}}}
\newcommand{\cdm}{\mathrm{cdm}}
\newcommand{\idm}{\mathrm{idm}}
\newcommand{\dr}{\mathrm{dr}}

\newcommand{\Neff}{\ensuremath{N_{\mathrm{eff}}}}
\newcommand{\DNeff}{\ensuremath{\Delta N_\mathrm{eff}}}
\newcommand{\NIR}{\ensuremath{\DNeff^\text{IR}}}
\newcommand{\NUV}{\ensuremath{\DNeff^\text{UV}}}
\newcommand{\HO}{\ensuremath{H_0}}
\newcommand{\Se}{\ensuremath{S_8}}

\newcommand{\vx}{\vec x}
\newcommand{\vk}{\vec k}
\newcommand{\vp}{\vec p}

\begin{document}
\preprint{\begin{flushright}
    UTTG-10-2022\\
\end{flushright}}

\title{Stepped Partially Acoustic Dark Matter, Large Scale Structure, and the Hubble Tension}

\date{\today}
\author[a,c]{Manuel A. Buen-Abad,}
\author[a]{Zackaria Chacko,}
\author[b]{Can Kilic,}
\author[a]{Gustavo Marques-Tavares,}
\author[b]{Taewook Youn}

\affiliation[a]{Maryland Center for Fundamental Physics, Department of Physics, University of Maryland, College Park, MD 20742, U.S.A.}
\affiliation[b]{Center for Theory, Weinberg Institute, Department of Physics, University of Texas at Austin, Austin, TX 78712, U.S.A.}
\affiliation[c]{Dual CP Institute of High Energy Physics, C.P. 28045, Colima, M\'{e}xico}
\emailAdd{buenabad@umd.edu}
\emailAdd{zchacko@umd.edu}
\emailAdd{gusmt@umd.edu}
\emailAdd{kilic@physics.utexas.edu}
\emailAdd{taewook.youn@utexas.edu}

\abstract{

We propose a new interacting dark sector model, Stepped Partially Acoustic Dark Matter (SPartAcous), that can simultaneously address the two most important tensions in current cosmological data, the $H_0$ and $S_8$ problems. As in the Partially Acoustic Dark Matter (PAcDM) scenario, this model features a subcomponent of dark matter that interacts with dark radiation at high temperatures, suppressing the growth of structure at small scales and thereby addressing the $S_8$ problem. However, in the SPartAcous model, the dark radiation includes a component with a light mass that becomes non-relativistic close to the time of matter--radiation equality. As this light component annihilates away, the remaining dark radiation heats up and its interactions with dark matter decouple. The heating up of the dark sector results in a step-like increase in the relative energy density in dark radiation, significantly reducing the $H_0$ tension, while the decoupling of dark matter and dark radiation ensures that the power spectrum at larger scales is identical to \lcdm.

}

\maketitle


\section{Introduction}
\label{sec:intro}

The standard cosmological model, \lcdm, has been remarkably successful in describing the increasingly precise observations of the past few decades \cite{Peebles:1984ge, Carroll:2000fy, Peebles:2002gy}. However, in recent years, there has been increasing tension between the determinations, using different datasets and methods, of two parameters: $\HO$, which parameterizes the current expansion rate of the universe, and $\Se$ (or the related $\sigma_8$), which describes the matter power spectrum at scales around $8 h^{-1}$ Mpc. In the case of $\HO$, measurements that rely on a \lcdm~fit to cosmological data, such as the cosmic microwave background (CMB) and the matter power spectrum, favor lower values of $\HO \approx 68~\HOunit$~\cite{Planck:2018vyg, ACT:2020gnv, SPT-3G:2021eoc, Schoneberg:2019wmt, Addison:2013haa, Aubourg:2014yra, Addison:2017fdm, Blomqvist:2019rah, Cuceu:2019for, Verde:2016ccp, Bernal:2021yli}, while more direct determinations based on standard candles and distance measurements favor larger values, $\HO \gtrsim 70~\HOunit$~\cite{Riess:2021jrx, Freedman:2019jwv, Freedman:2021ahq, Yuan:2019npk, Soltis:2020gpl, Khetan:2020hmh, Huang:2019yhh, Schombert:2020pxm, Wong:2019kwg, Birrer:2020tax, Pesce:2020xfe, LIGOScientific:2019zcs}. Comparing the most precise measurements in each of these two categories, namely the \lcdm~fit to Planck CMB data and the supernovae measurements made by the SH0ES collaboration calibrated to Cepheid variable stars~\cite{Riess:2021jrx}, the tension has reached $5\sigma$ significance~\cite{Verde:2019ivm, DiValentino:2021izs, Schoneberg:2021qvd}. The tension between $\Se$ determinations, while not as drastic as that for $\HO$, arises from the fact that direct $\Se$ measurements using weak lensing, cluster counts, or galaxy power spectra have yielded consistently lower values than the determination of $\Se$ from fits of \lcdm~parameters to Planck data~\cite{Planck:2018vyg, Joudaki:2016mvz, HSC:2018mrq, Heymans:2020gsg, DES:2021wwk, KiDS:2021opn, White:2021yvw}. While it is possible that one or both of these tensions arise from unknown systematic effects, if confirmed, they point to the necessity of introducing new physics beyond \lcdm.

The $\HO$ tension has received a great deal of attention and many theories have been put forward to address it~(see Refs.~\cite{DiValentino:2021izs, Schoneberg:2021qvd, Abdalla:2022yfr} and references therein for a comprehensive list of proposals). One class of approaches to the problem, the so-called ``late universe" solutions~(see e.g.~\cite{Zhao:2017cud,DiValentino:2017gzb,Krishnan:2020vaf,Dainotti:2021pqg,Odintsov:2022eqm,Colgain:2022rxy}), rely on modifications to the late-time expansion of the Universe, or on new physics impacting the cosmic ladder measurements of $\HO$. The alternative class of approaches, the ``early universe" solutions, are instead based on modifications to the cosmic history at early times.  Since there are strong indications that addressing the $\HO$ tension requires a decrease in the sound horizon~\cite{Bernal:2016gxb,Aylor:2018drw,Knox:2019rjx}, i.e. the distance traveled by sound waves in the photon-baryon plasma prior to recombination, this is the path we shall follow. One of the most promising ways to realize this is by having an extra contribution to the energy density around the time of matter–radiation equality. This extra energy density increases the Hubble expansion rate during that epoch, leading to a smaller sound horizon.

In order to remain consistent with CMB and large-scale structure (LSS) data, the new contribution to the energy density must redshift faster than matter after matter--radiation equality. Perhaps the simplest way to realize this is by introducing into the theory a new dark radiation (DR) component. This takes the form of one or more new relativistic species that have non-negligible energy densities at the relevant times. DR that behaves like massless neutrinos (\ie that has negligible self-interactions) only provides a mild improvement due to constraints from CMB~\cite{Bernal:2016gxb, Planck:2018vyg}. While adding self-interactions to the DR does help, it does not fully resolve the $\HO$ tension~\cite{Baumann:2015rya, Brust:2017nmv, Blinov:2020hmc} (see also Refs.~\cite{Kreisch:2019yzn, Escudero:2019gvw, RoyChoudhury:2020dmd, Brinckmann:2020bcn} for scenarios in which all or part of the DR is initially interacting and transitions to free streaming before recombination).

In recent work \cite{Aloni:2021eaq}, it was shown that a simple modification of the DR scenario, dubbed StepDR by the authors, can lead to a significant reduction in the Hubble tension. In this construction, the DR, which is self-interacting, consists of two components. One of the components has a mass of order eV, while the other component is massless. Then, once the temperature in the dark sector falls below the mass of the massive species, it annihilates away into the massless component, which heats up. This results in a step-like increase in the relative energy density in DR around the redshift at which the massive species becomes non-relativistic. If this step happens shortly before matter--radiation equality, it can provide an improved fit to the CMB and a compelling solution to the $\HO$ tension. Remarkably, a minimal realization of the StepDR framework, in which the role of the DR is played by a supersymmetric Wess-Zumino model, can provide a good fit to the data. This specific realization of StepDR is referred to as Wess-Zumino Dark Radiation (WZDR) by the authors of Ref.~\cite{Aloni:2021eaq}.

Another type of energy density that can also provide a significant reduction in the tension is the contribution from a new scalar field, whose energy density behaves as a cosmological constant at early times but dilutes away rapidly around the time of matter--radiation equality~\cite{Karwal:2016vyq,Poulin:2018cxd} (see also~\cite{Alexander:2019rsc,Berghaus:2019cls,Ye:2020btb,Niedermann:2021vgd,Ye:2021iwa,Berghaus:2022cwf}). In this class of models, dubbed Early Dark Energy (EDE), the degree to which the tension is lessened depends sensitively on the form of the scalar field potential. However, it is pointed out in ~\cite{Agrawal:2019lmo} that the simplest potentials only lead to a modest reduction in the tension. Although more exotic forms of the potential can provide a better fit~\cite{Smith:2019ihp,Lin:2019qug}, constructing ultraviolet-complete models that realize these potentials remains a challenge.

Addressing the $\Se$ tension requires a decrease in the matter power spectrum on scales with $k \sim 0.1 - 1 h/\Mpc$ with respect to $\Lambda$CDM. Some of the early proposals suggested that neutrino masses and the addition of sterile neutrinos could achieve this, but were disfavored when fit to data~(see e.g. \cite{Leistedt:2014sia,Battye:2014qga}). A more promising approach has been to evoke new dynamics for dark matter (DM), such as interactions with a dark sector~\cite{Buen-Abad:2015ova,Lesgourgues:2015wza,Murgia:2016ccp,Kumar:2016zpg,Chacko:2016kgg,Buen-Abad:2017gxg,Buen-Abad:2018mas,Dessert:2018khu,Archidiacono:2019wdp,Heimersheim:2020aoc,Bansal:2021dfh} or DM decays~\cite{Enqvist:2015ara,Poulin:2016nat,Clark:2020miy,FrancoAbellan:2021sxk}. It is important to note that most models that aim to solve the $\HO$ tension by increasing the energy density around matter--radiation equality exacerbate the $\Se$ tension. This is the case in StepDR, due to the presence of the extra DR. Solutions to the $\HO$ tension based on EDE face similar challenges, with a joint analysis of CMB and LSS data appearing to disfavor this scenario~\cite{Ivanov:2020ril,DAmico:2020ods} (although the robustness of these claims has been disputed~\cite{Smith:2020rxx}). This suggests that the solutions to the $\HO$ and $\Se$ tensions might be intertwined and require a joint resolution.

Among the solutions that have been put forward to resolve the $\Se$ problem, there are two that give rise to a modest relaxation of the $\HO$ tension as well, and are therefore of particular interest. These are the Non-Abelian Dark Matter Dark Radiation (NADMDR) model proposed in \cite{Buen-Abad:2015ova, Lesgourgues:2015wza} and the Partially Acoustic Dark Matter (PAcDM) model proposed in \cite{Chacko:2016kgg}. In both of these constructions there is a self-interacting DR component that also has interactions with DM. In the NADMDR framework, all of DM interacts with DR, but these interactions are very weak and never come into thermal equilibrium. In the PAcDM scenario, only a subcomponent of DM interacts with DR, but these interactions can be much larger and are in thermal equilibrium at early times. In both the NADMDR and the PAcDM models, the effect of the DR--DM interactions is to slow down the rate of structure formation, thereby providing a solution to the $\Se$ tension. In addition, the presence of interacting DR can somewhat ameliorate the $\HO$ tension. However, as already stated, self-interacting DR models {\it without a step} only lead to a modest improvement in the $\HO$ tension~\cite{Buen-Abad:2017gxg}.

In this work, we propose {\it ``Stepped Partially Acoustic Dark Matter''} (SPartAcous), a simple modification of the PAcDM model inspired by the StepDR scenario, which can address both the $\HO$ and $\Se$ problems. In the original PAcDM model, the DR consists of massless dark fermions and the gauge bosons of an unbroken $U(1)$ gauge symmetry. The fermions carry a dark charge under the $U(1)$, and are therefore coupled to the gauge bosons. A subcomponent of DM is also charged under the $U(1)$. The resulting interactions between DM and DR offer a solution to the $\Se$ problem. The generalization of PAcDM to SPartAcous simply involves introducing a small mass term for the dark fermions. Then, once the dark sector temperature drops below the fermion mass, the dark fermions annihilate away into gauge bosons. This dark sector mass threshold induces two important effects. Firstly, it introduces a step in the relative energy density in DR, just as in the original StepDR proposal. Secondly, it induces a change in the dynamics of the charged DM sub-component by turning off its pressure and allowing its perturbations to grow. Consequently, the transition from partially acoustic to collisionless DM tends to occur earlier in SPartAcous than in PAcDM, allowing a better fit to the CMB.

In this paper, we motivate and introduce the SPartAcous framework, present the underlying particle physics model, and determine its impact on various cosmological observables. The results in this paper have been obtained using a modified version of the cosmological code {\tt CLASS} \cite{class}. In upcoming work~\cite{Buen-Abad:2023b}, we will perform a dedicated likelihood analysis of the SPartAcous model to the available cosmological data.

This paper is organized as follows. In Sec.~\ref{sec:model}, we introduce the model and highlight its most important features. In Sec.~\ref{sec:cosmo}, we study its effects on cosmological evolution and determine the evolution of cosmological perturbations in the dark sector. In Sec.~\ref{sec:results}, we present numerical results that showcase the characteristic features of the model, and present evidence that it can significantly improve both cosmological tensions. In the final, section we conclude. Some details of the analysis are presented in the appendix.

\section{The Model}
\label{sec:model}

In this section, we present the SPartAcous model, describe its field content and interactions, and explain how these give rise to the desired cosmological history. The SPartAcous model is characterized by an interacting dark sector which, in addition to DM, also contains DR. As in PAcDM, here DM is composed of two distinct components: a primary component that is cold and non-interacting, and a subdominant component that has sizable interactions with the DR. The role of the subdominant component is played by a complex scalar field $\chi$ that carries charge under a dark $U(1)$ gauge symmetry.  Going forward, $\chi$ will be referred to as the interacting DM (iDM) component\footnote{The role of iDM could also be played by a charged fermion without changing the main features of the model. Our choice of a scalar is simply to match the original PAcDM model~\cite{Chacko:2016kgg}.}. The massless gauge boson associated with the gauge symmetry is a component of the DR, and is labelled by $A_\mu$. The dark sector also contains a light Dirac fermion $\psi$ with mass $m_\psi$ that is charged under the $U(1)$ gauge symmetry. Then the Lagrangian for the dark sector takes the form,
\begin{equation}
\label{eq:lagrangian}
    \delta \Lag_\text{dark} = -\frac{1}{4} V_{\mu\nu}V^{\mu\nu} + \bar \psi (i \slash  \!\!\!\! D - m_\psi) \psi + |D \chi|^2 - m_\chi^2 |\chi|^2\, .
\end{equation}
Here $V_{\mu\nu}$ is the field strength associated with the dark $U(1)$ gauge field, and $D_\mu$ is the associated covariant derivative. For simplicity we have assumed that $\chi$ and $\psi$ have the same charges under the $U(1)$. The generalization to the case with arbitrary charges is straightforward. We will be considering $\chi$ masses of order the weak scale. There are no significant cosmological constraints from self-interactions for DM in this mass range, and so while a scalar quartic coupling can be included for completeness, it has no significant impact on our results. We assume the dark sector has its own temperature $T_d$, which is slightly lower than the visible sector temperature $T$. This is quite natural, since even if the two sectors are in equilibrium at very early times, after they decouple the SM bath will get heated up as the heavy species annihilate away. 

This scenario can be embedded in a larger dark sector, as in the PAcDM proposal~\cite{Chacko:2016kgg}, in which there are extra fields that can play the role of the dominant cold DM (CDM) component, and additional interactions that can explain the cosmic abundances of iDM and CDM. The iDM component, $\chi$, is assumed to have a cosmological abundance that amounts to a small fraction of the total DM density, $\omega_\chi/\omega_{\rm DM} \sim 10^{-2}$. This can easily be arranged within the context of the standard freeze-out framework, simply by allowing for different DM masses and correspondingly different DM annihilation cross sections, see Refs.~\cite{Chacko:2015noa, Chacko:2016kgg}. However, the associated dynamics is relevant only at very large dark sector temperatures ($T_d \gg \mathrm{keV}$), and is therefore not directly probed by CMB or LSS experiments. For this reason, in this work we will remain agnostic about the physics at very high temperatures that explains the cosmic abundances of the various particles in the dark sector, and limit our attention to the dynamics that happens after the temperature has dropped below $\sim \keV$, after which it can start impacting the relevant observables.

\begin{figure}[tb]
	\centering
	\includegraphics[width=.4\linewidth]{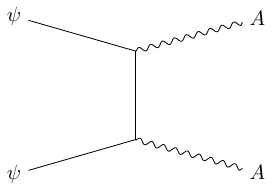}
	\includegraphics[width=.4\linewidth]{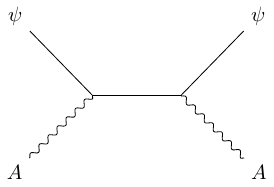}
	\includegraphics[width=.4\linewidth]{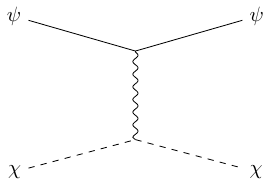}
	\includegraphics[width=.4\linewidth]{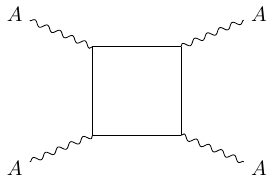}
	\caption{Feynman diagrams for the relevant dark sector interactions. {\bf Upper Left:} Pair annihilation/creation of $\psi$ into/from dark gauge bosons $A$. {\bf Upper Right:} Compton scattering between $\psi$ and $A$, which keeps the two fluids tightly coupled. {\bf Lower Left:} Long-range interactions between $\psi$ and the iDM component $\chi$ through the exchange of dark gauge bosons. {\bf Lower Right:} Self-interaction of $A$ from the Euler-Heisenberg operator, with $\psi$ running in the loop. The upper diagrams play a role in keeping the DR in equilibrium at high temperatures, and the lower right diagram does the same at low temperatures. The lower left diagram makes DR-iDM interactions possible.}
	\label{fig:diagrams}
\end{figure}

The processes that play an important role in the cosmological history are illustrated in the Feynman diagrams of \Fig{fig:diagrams}. At high temperatures, $T_d \gtrsim m_\psi$, the dark fermion and gauge boson behave as DR. The upper left diagram shows the pair annihilation of $\psi$ into gauge bosons. The cross section for this process is of the order of,
\begin{equation}\label{eq:dr-annihilation}
\langle \sigma_{\psi\,\text{ann}} v \rangle \sim \left\{  
	\begin{array}{lc}
		\frac{\alpha_d^2}{T_d^2} \, , & T_d > m_\psi \, , \\
		\frac{\alpha_d^2}{m_\psi^2} \, , & T_d < m_\psi \, ,
	\end{array}	
\right.
\end{equation}
with $\alpha_d \equiv g_d^2/(4\pi)$. At temperatures $T_d \gg m_\psi$, pair creation also happens at the same rate as pair annihilation.

The upper right diagram shows the Compton scattering of $\psi$ and $A$, which at high temperatures has a cross section approximately given by
\begin{equation}
\label{eq:dr-dr-scattering}
	\langle \sigma_{\psi A} v \rangle \sim \frac{\alpha_d^2}{T_d^2} \, .
\end{equation}
Together, pair annihilation, pair creation and Compton scattering keep $\psi$ and $A$ in thermal equilibrium at high temperatures, $T_d > m_\psi$. At these temperatures, the dark fermion and gauge boson constitute interacting DR.

Since we know the number of relativistic degrees of freedom in the dark sector at these temperatures, we can relate $T_d$ to the energy density in DR. This is conventionally parameterized in units of the energy density in a single neutrino species, $\DNeff$. At early times, when the temperature in the dark sector is well below the electron mass $m_e$ but well above the mass of $\psi$, so that $m_e \gg T_d \gg m_\psi$, the ratio of the temperatures in the visible sector and dark sectors remains constant. The contribution of DR to $\DNeff$ in this regime is given by
\begin{equation}
\NUV = \frac{\rho_A + \rho_\psi}{\rho_{1\nu}} = \frac{\left(2 + \frac{7}{2}\right)}{\frac{7}{4} \left( \frac{4}{11} \right)^{4/3}} \,
\bl( \frac{T_d}{T} \br)^4 \approx 12.1 \, \bl( \frac{T_d}{T} \br)^4 \, .
\end{equation}
Here $\rho_A$ and $\rho_\psi$ denote the energy densities in dark gauge bosons and dark fermions respectively, and $\rho_{1 \nu}$ represents the energy density in a single neutrino species. The last equality assumes $T \ll m_e$. From this we see that in a scenario with $\DNeff \sim 0.1 - 1$, the range which can potentially address the $\HO$ problem~\cite{Aloni:2021eaq}, the temperature of the dark sector cannot be very different from that of the visible sector.

The lower left diagram shows the dominant interaction between iDM and DR, which at temperatures below $m_\chi$ primarily arises from the $t$-channel exchange of dark gauge bosons between $\psi$ and $\chi$. Just as in conventional Rutherford scattering, this cross section is technically divergent. However, as we will see in the next section, the relevant interaction rates are those of momentum exchange and heat exchange, which are finite once plasma effects are taken into account. This process keeps the iDM and DR in kinetic equilibrium at high temperatures, $T_d > m_\psi$. This suppresses the growth of structure for modes that enter the horizon at these early times, thereby offering a solution to the $\Se$ problem.
	
In principle, Compton scattering between $A$ and $\chi$ (same as the upper right diagram of \Fig{fig:diagrams} but with $\psi$ replaced by $\chi$) could also be relevant and lead to an additional coupling between iDM and DR. The cross section for this process is of the order of
\begin{equation}
\langle \sigma_{\chi A} v \rangle \sim \frac{\alpha_d^2}{m_\chi^2} \, .
\end{equation}
However, as long as $m_\chi$ is large enough, this interaction will be sufficiently feeble at low temperatures so as to not play an important role. 

At temperatures below the mass of $\psi$, the dark fermions annihilate away into gauge bosons through the diagram on the top left of Fig.~\ref{fig:diagrams}, and their number density becomes exponentially suppressed. The DR bath, now consisting only of gauge bosons, gets heated up. This generates a step-like increase in $\DNeff$, with the low-temperature value being larger by a factor that depends on the ratio of high-to-low temperature effective number of degrees of freedom in the DR. This is analogous to what happens to the photon temperature after electrons and positrons become non-relativistic and annihilate away. The high temperature and low temperature values of $\Neff$ are related by,
\beq
    \NIR = \bl( \frac{11}{4} \br)^{1/3} \, \NUV \approx 1.4 \, \NUV \ .
\eeq
The presence of this step feature in $\Neff$ allows a solution to the $\HO$ tension along the lines of Ref.~\cite{Aloni:2021eaq}. We present a more detailed study of the redshift evolution of the dark sector temperature and $\DNeff$ in \App{A3-bg}.

Once the dark fermions have exited the bath, the process shown in the bottom left diagram of \Fig{fig:diagrams} is no longer effective, and so iDM and DR are no longer in kinetic equilibrium. Consequently, the long wavelength modes that entered the horizon after $\psi$ exited the bath exhibit no suppression in structure relative to $\Lambda$CDM.

Even though the DR bath now consists only of gauge bosons, it can still have sizable self-interactions through the loop diagram shown in the lower right of Fig.~\ref{fig:diagrams}. If the energies involved in the $AA\rightarrow AA$ scattering are much smaller than $m_\psi$, the effects of this diagram can be approximated using the effective Euler-Heisenberg type operator
\begin{equation}
\mathcal{L}_D^\mathrm{EH} = \frac{\alpha_d^2}{90 m_\psi^4} \left[ (V_{\mu\nu}V^{\mu\nu})^2 + \frac{7}{4} (V_{\mu\nu}\tilde V^{\mu\nu})^2 \right] \, .
\end{equation}
In this limit, the cross section is approximately given by
\begin{equation}
\label{eq:self-scattering}
        \langle \sigma_{AA} v \rangle \sim \alpha_d^4 \frac{T^6}{m_\psi^8} \, .
\end{equation}
This scattering process allows the DR to continue to be self-interacting even at temperatures below the mass of $\psi$.

We see from this discussion that the SPartAcous model has all the features necessary to address both the $\Se$ and $\HO$ problems. We will discuss the cosmological history in greater detail in the next section.

\section{Cosmological History}
\label{sec:cosmo}

\subsection{Background Evolution}
\label{subsec:bg}

In this subsection we consider how the cosmological parameters evolve in the SPartAcous model at the background level, and determine the range of parameter space for which the scenario is viable. We will consider perturbations about the background in the next subsection. 

We begin by determining the region of parameter space in which at high temperatures, $T_{d} > m_\psi$, the components of the DR are in thermal equilibrium. This requires the scattering and annihilation rates, shown in Eqs.~(\ref{eq:dr-annihilation}) and~(\ref{eq:dr-dr-scattering}), be larger than the Hubble expansion rate. In terms of the coupling constant $\alpha_d$ this requires
\begin{equation}
\label{eq:DR-2-to-2}
 \alpha_d^2 \, T_d \gtrsim \frac{T^2}{M_p} \Rightarrow \alpha_d \gtrsim 10^{-12} \, \bl( \frac{T}{1~\keV} \br)^{1/2} \, ,
\end{equation}
where the largest relevant temperature $T\sim \keV$ leads to the strongest constraint. In this estimate, and also similar future ones, we have taken $T_d \sim T$, which suffices for $O(1)$ comparisons. Note also that the dark photon chemical potential vanishes as long as the rates for 2-to-3 processes, such as $\psi A \rightarrow \psi A A$, are larger than the expansion rate. This corresponds to the following condition on $\alpha_d$,
\begin{equation}
	\alpha_d^3 \, T_d \gtrsim \frac{T^2}{M_p} \Rightarrow \alpha_d \gtrsim 10^{-8} \bl( \frac{T}{1~\keV} \br)^{1/3} \, .
\end{equation}

Once the dark sector temperature falls below $m_\psi$, the dark fermions begin to go out of the bath. The number density of $\psi$ freezes out when
\begin{equation}
 \frac{m_\psi}{T_d} \sim 39 + \log \left( \frac{\alpha_d}{10^{-5}}\right)^2 \, ,
\end{equation}
leaving a small relic abundance of $\psi$,
\begin{equation}
 \frac{\Omega_\psi}{\Omega_\text{CDM}} \sim \frac{10^{-9} \, \text{GeV}^{-2}}{\alpha_d^2/m_\psi^2} \sim 10^{-17} \,
\left( \frac{m_\psi}{\text{eV}} \right)^2  \left( \frac{10^{-5}}{\alpha_d} \right)^2 \, .
\end{equation}
For the typical parameters of interest, the relic abundance of $\psi$ after it has frozen out is negligible, and it does not play a role in cosmological evolution beyond this point. Therefore, its precise value is irrelevant for our results. During the freeze out process, the dark fermions remain thermally coupled to the gauge bosons through Compton scattering.

In our numerical study, we will treat the DR as a perfect fluid. Accordingly, we require that even below the mass of $\psi$, the DR bath remains self-interacting at least until recombination ($T \sim 0.1 \, \eV$), after which its contribution to the energy density is small. From Eq.~(\ref{eq:self-scattering}), we find that the self-interactions decouple when
\begin{equation}
 \Gamma_{AA} \sim \frac{\alpha_d^4}{m_\psi^8} T_d^9 \sim \left( \frac{T_\eq}{1~\eV} \right)^{1/2} \frac{T^{3/2}}{M_p} \  \Rightarrow \ T_{A\,\dec} \sim 0.1~\eV~ \bl( \frac{m_\psi}{\eV} \br)^{\frac{16}{15}} 
\bl( \frac{10^{-5}}{\alpha_d} \br)^{\frac{8}{15}} \ ,
\end{equation}
where $T_\eq$ is the temperature at matter--radiation equality, and where we assumed that the decoupling happens for $T < T_\eq$. We see from this that, provided $\alpha_d \gtrsim 10^{-5}$, the DR can be treated as a tightly coupled fluid during the relevant times of its cosmological evolution.

The heating rate $\dot Q_{\psi \chi}$ quantifies the rate of heat transfer from the dark fermions to the DM~\cite{Bringmann:2006mu,Cyr-Racine:2015ihg}~(see also Ref.~\cite{Ma:1995ey} where this quantity is referred to as the specific heating rate in the context of baryons and photons). The iDM remains kinetically coupled to the DR at high temperatures (through the lower left diagram in \Fig{fig:diagrams}) as long as the heating rate from $\psi$ to $\chi$ satisfies the condition,
\begin{equation}
 \label{eq:Qchipsi}
	\dot Q_{\psi \chi} \sim \frac{\alpha_d^2 T_d^2}{m_\chi} \gtrsim \frac{T^2}{M_p} \Rightarrow \alpha_d \gtrsim 10^{-8} \, \bl( \frac{m_\chi}{10^3~\GeV} \br)^{1/2}.
\end{equation}

In order to ensure that the interactions between DR and iDM turn off below the mass threshold, the direct scattering rate between the gauge bosons and $\chi$ must be sufficiently small. This requires that the heating rate from $A$ to $\chi$ (and also the related momentum-exchange rate) is small compared to the expansion rate at temperatures below the mass of the dark fermion,
\begin{equation}
	\dot Q_{\chi A} \sim \frac{\alpha_d^2 T_d^4}{m_\chi^3} \lesssim \frac{T^2}{M_p} \Rightarrow \alpha_d \lesssim 10 ~\left(\frac{m_\chi}{10^3 ~ \GeV}\right)^{3/2} \bl( \frac{1~\keV}{T} \br) \, .
\end{equation}
A conservative choice is to require that this condition is already true when $T = \keV$, which is much earlier than the modes constrained by CMB and $\Se$ have entered the horizon.

Lastly, the iDM annihilation rate into DR cannot be so large that the freeze-out abundance one obtains by considering that process alone leads to a smaller $f_\chi = \rho_\chi/\rho_\text{DM}$ than what is needed to address the $\Se$ problem. As long as the annihilation rate through $A$ is sufficiently small, there can be other interactions that are important at early times, $T\sim m_\chi$, leading to the correct $f_\chi$. Requiring that processes mediated by $A$ alone lead to an abundance greater than a fraction $f_\chi$ of
the observed dark matter abundance, we obtain 
\begin{equation}
\label{eq:dark-abundance}
 f_\chi \lesssim \frac{10^{-9} \, \text{GeV}^{-2}}{\alpha_d^2/m_\chi^2} 
 \sim 10 \bl( \frac{m_\chi}{\GeV} \br)^2 
\bl( \frac{10^{-5}}{\alpha_d} \br)^2 
\, .
\end{equation}
 In this estimate we have ignored the effects of Sommerfeld enhancement associated with the long range interactions mediated by $A$. The Sommerfeld enhancement~\cite{Kamionkowski:2008gj, Arkani-Hamed:2008hhe}, in which the annihilation cross-section is enhanced at low velocities by $1/v$, can lead to a $\mathcal{O}(10)$ impact on the abundance for $\alpha_d \gtrsim 0.1$, but is negligible for $\alpha_d \lesssim 10^{-2}$ \cite{Dent:2009bv}.

\begin{figure}[t]
	\centering
	\includegraphics[width=.6\linewidth]{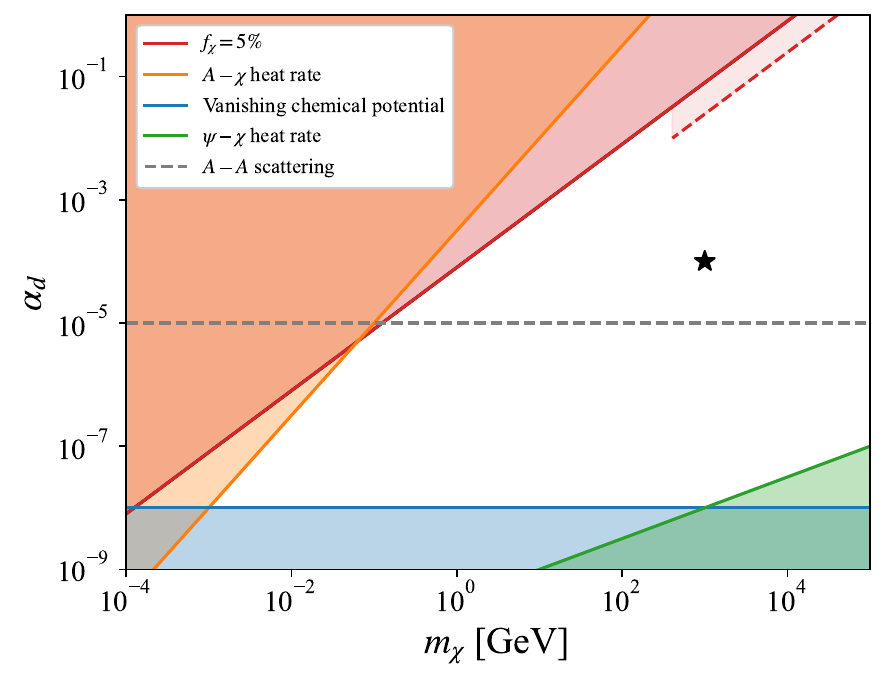}
	\caption{Viable parameter space satisfying all the conditions discussed in Sec.~\ref{subsec:bg}, with $m_\psi$ set to $1~\eV$. The vanishing chemical potential and the $\psi - A$ heating constraint are imposed at $T = \keV$. The annihilation constraint requires that the abundance of iDM is consistent with $f_\chi = 5\%$. The red dashed line shows the annihilation constraint taking into account Sommerfeld enhancement, which is only relevant for $\alpha_d \gtrsim 10^{-2}$. This limit assumes that Sommerfeld enhancement is sufficiently large for s-channel annihilation at freeze-out to grow as $1/v$, although is strictly only a good approximation for $\alpha_d \gtrsim 0.1$~\cite{Dent:2009bv}. The region above the horizontal gray dashed line corresponds to those parameters for which the $A$--$A$ scattering through $\psi$ loops keep the DR self-interacting until the time of recombination. This allows the interacting radiation approximation used in Sec.~\ref{subsec:pert} to be valid at all relevant times. The star, at $m_\chi = 10^3 ~\GeV$ and $\alpha_d = 10^{-4}$, marks the benchmark point that we will use in \Sec{sec:results}.}
	\label{fig:param-space}
\end{figure}

In \Fig{fig:param-space}, we illustrate the allowed $(\alpha_d, \, m_\chi)$ parameter space in which all the requirements listed above are satisfied. We have fixed $m_\psi = 1 \, \eV$, which corresponds to the step in $\Neff$ taking place around the time of matter--radiation equality (as seems to be preferred by the data \cite{Aloni:2021eaq}). In the constraint for annihilation, arising from Eq.~(\ref{eq:dark-abundance}), we required that the annihilations are consistent with $f_\chi = 5\%$, since we will be interested in $f_\chi \lesssim 5\%$. We see that all of the assumptions in this section are satisfied for a wide range of dark sector parameters, showing that the scenario we are interested in is fairly generic in interacting dark sector models with a mass threshold. For future reference, we introduce a benchmark parameter point at $m_\chi = 10^3~\GeV$ and $\alpha_d = 10^{-4}$.

\subsection{Evolution of Perturbations}
\label{subsec:pert}

In this subsection, we study the evolution of perturbations in the dark sector, following closely the notation in Ref.~\cite{Ma:1995ey}. As discussed in the previous section, we will be interested in a gauge coupling sufficiently large that the DR behaves as a self-interacting perfect fluid until at least recombination. After that point the impact of the DR on the relevant observables is very small, and so whether the DR is self-interacting or not becomes immaterial.

Under this assumption, we can treat $\psi$ and $A$ as a single perfect fluid for the entire time period relevant to our analysis. For perfect fluids the perturbations of the DR are described in terms of their local density and velocity divergence perturbations, $\delta_\dr$ and $\theta_\dr$ respectively. We can write the equations for the perturbations in the DR and iDM in the conformal Newtonian gauge,
\begin{eqnarray}
	\dot \delta_\idm & = &  - \theta_\idm + 3 \dot \phi  \, , \\
	\dot \theta_\idm & = & - \mH \theta_\idm + k^2 \psi + a \Gamma \left( \theta_\dr - \theta_\idm \right) \, , \\
	\dot \delta_\dr & = &  - (1+w)(\theta_\dr - 3 \dot \phi) - 3 \mH \left(c_s^2 - w \right) \delta_\dr \, ,\\
	\dot \theta_\dr & = & - \left[ (1-3w) \mH  + \frac{\dot w}{1+w} \right] \theta_\dr + k^2 \left(\frac{c_s^2}{1+w}  \delta_\dr +\psi \right) \nonumber\\
	& & + \frac{\rho_\idm}{\rho_\dr (1+w)} \, a \Gamma (\theta_\idm - \theta_\dr) \ .
\end{eqnarray}
Here the dot denotes derivatives with respect to conformal time, $a$ is the scale factor of the Universe, $\mH \equiv a H = \dot a/a$ is the conformal Hubble expansion rate, $w$ and $c_s$ are the equation of state and the sound speed for the DR respectively (which we study in \App{A3-bg}), and $\Gamma$ is the momentum-exchange rate between iDM and DR~\cite{Buen-Abad:2015ova,Cyr-Racine:2015ihg,Buen-Abad:2017gxg}. For a derivation of these equations, see \App{A4-perts}. In the limit of interest, when $T_d \ll m_\chi$, the interaction rate is given by (see \App{A5-rate}),
\beq
\label{eq:momentum-transfer}
    \Gamma = \frac{4}{3 \pi} \, \alpha_d^2 \, \ln (4/\langle \theta_{\rm min} \rangle^2) \, \frac{T_d^2}{m_\chi} \, e^{-m_\psi/T_d}\left[2 + \frac{m_\psi}{T_d}\left(2+\frac{m_\psi}{T_d}\right)\right] \ .
\eeq
In obtaining this expression we have regularized the divergence in forward scattering by imposing a minimum scattering angle $\theta_{\rm min}$. This angle is set by the Debye screening length of the DR. For more details, see \App{A5-rate}. For our benchmark choices of $\alpha_d$ and $m_\chi$, the ratio of $\Gamma$ and the Hubble rate $H$ in the ultraviolet ($T_d \sim T \gg m_\psi$) is roughly given by:
\beq
    \frac{\Gamma}{H} \sim 10^9 ~ \bl( \frac{\alpha_d}{10^{-4}} \br)^2 \bl( \frac{10^3~\GeV}{m_\chi} \br) \ ,
\eeq
showing that we are well within the iDM--DR tightly coupled regime.

The very large interaction rate between iDM and DR prevents the growth of perturbations in the iDM fluid, analogous to the situation in the PAcDM model \cite{Chacko:2016kgg}. This slows the growth of DM perturbations since the iDM component contributes to the expansion rate like a matter field but does not gravitationally cluster. This changes the evolution of the gravitational potential perturbations and slows the growth of the dominant CDM fluid.

Once $T_d < m_\psi$, the interaction rate between the DR and iDM becomes exponentially suppressed due to exponential suppression of the $\psi$ number density, and quickly becomes irrelevant. At this point, there are two important differences between SPartAcous and the PAcDM scenario. Firstly, the perturbations entering the horizon after this decoupling will behave as CDM, and so the resulting power spectrum at long wavelengths is the same as in a scenario in which all of DM is standard CDM. In PAcDM, the transition to CDM like behavior only occurs when the energy density in iDM exceeds that in DR~\cite{Buen-Abad:2017gxg}. This implies that for large $\DNeff$, the transition tends to occur well into the regime that is significantly constrained by CMB. Secondly, since the iDM completely decouples from the DR, its sound speed quickly becomes negligible. This means that in SPartAcous, even those iDM perturbations that entered the horizon {\it before} this decoupling grow quickly and catch up with the perturbations in the CDM fluid. {\it Mutatis mutandis}, a similar behavior takes place in the baryon perturbations after they decouple from the photons. Therefore, shortly after decoupling, the DM sector, comprised of both CDM and iDM, behaves as in \lcdm, but with a non-trivial primordial power spectrum at short wavelengths compared to \lcdm~models. In addition, a ``memento'' of the iDM's tightly-coupled era is imprinted on the matter power spectrum once their perturbations catch up with those of the CDM. This takes the form of acoustic oscillations in the matter power spectrum, analogous to those from the baryons in the standard \lcdm~model (but at smaller scales for the parameter range we will be interested in). Following established convention, we refer to this phenomenon as {\it ``dark acoustic oscillations''} (DAO).

\section{Results}
\label{sec:results}

The results of this section have been obtained using the {\tt CLASS} code \cite{class}, which we have modified to implement the evolution of the DR and iDM background and perturbations as described in the previous section. The input parameters to the model are the low temperature contribution to $\DNeff$ after the transition, $\NIR$; the fraction of DM that is interacting $f_\chi$; and the redshift $z_t$ around which the step takes place (which, following Ref.~\cite{Aloni:2021eaq}, we define in terms of the transition scale factor, $(1+z_t)^{-1} \equiv a_t \equiv T_{d 0}/m_\psi$, where $T_{d0}$ is the DR temperature today). We use the benchmark point $m_\chi = 10^3 ~\GeV$ and $\alpha_d = 10^{-4}$ as previously mentioned. In the ultraviolet this translates to a tightly coupled system, with $\Gamma/H \sim 10^9$.

The WZDR model is quite successful in reducing the $\HO$ tension~\cite{Aloni:2021eaq}. Accordingly, we will focus on comparing the differences in the CMB predictions between our scenario and the best fit parameters of the WZDR model with respect to the dataset dubbed $\mathcal{D}+$ in Ref.~\cite{Aloni:2021eaq}. This dataset includes CMB observations from Planck \cite{Planck:2018vyg}, diverse BAO measurements \cite{BOSS:2016wmc, Beutler_2011, Ross:2014qpa}, the PANTHEON supernovae data \cite{Pan-STARRS1:2017jku}, and the $\HO$ determination from the SH0ES collaboration~\cite{Riess:2020fzl}.

Even though we fix $m_\chi$ and $\alpha_d$ for our numerical results in this section, it should be noted that when the DR and iDM fluids are in the tightly coupled limit, the exact values of $\alpha_d$ and $m_\chi$ (through the interaction rate $\Gamma$ in \Eq{eq:momentum-transfer}) are effectively irrelevant (up to small corrections of the order of $H/\Gamma$). Of course, once the dark sector temperature reaches the $\psi$ mass threshold, there is some sensitivity to these parameters, since the point at which the interactions decouple, \ie $\Gamma < H$, will depend on the combination $\alpha_d^2/m_\chi$. However, from Eq.~(\ref{eq:momentum-transfer}), we see that the rate drops exponentially with $m_\psi/T$, while it only depends linearly on $\alpha_d^2/m_\chi$. Therefore, the dependence on this combination of parameters is only logarithmic. For our benchmark values $\Gamma/H \sim 10^9$, which leads to this decoupling taking place roughly when $m_\psi / T_d \sim 20$.

We now proceed to explore the impact of SPartAcous on cosmological observables, in particular, $\HO$, $\Se$, the CMB, and the matter power spectrum. To better compare SPartAcous to WZDR, we fix our model's cosmological parameters to match the WZDR model's best fit values to the $\mathcal{D}+$ dataset, as found in Ref.~\cite{Aloni:2021eaq}. We do this with the exception of $\Omega_\text{DM} h^2$ and the SPartAcous parameters $\NIR$, $z_t$, $f_\chi$. We relate $\Omega_\text{DM} h^2$ to $\NIR$ by requiring that the redshift $z_\eq$ at matter--radiation equality be the same as at the best fit point of WZDR. Since $z_\eq$ is accurately measured by the Planck CMB observations, this is a well-motivated choice. We then compute the cosmological observables of interest for a grid of the SPartAcous parameters within the intervals $0.1 \leq \NIR \leq 1$, $0.5\% \leq f_\chi \leq 5\%$, and $3 \leq \log_{10} z_t \leq 5$.

\begin{figure}[t]
	\centering
	\includegraphics[width=.49\linewidth]{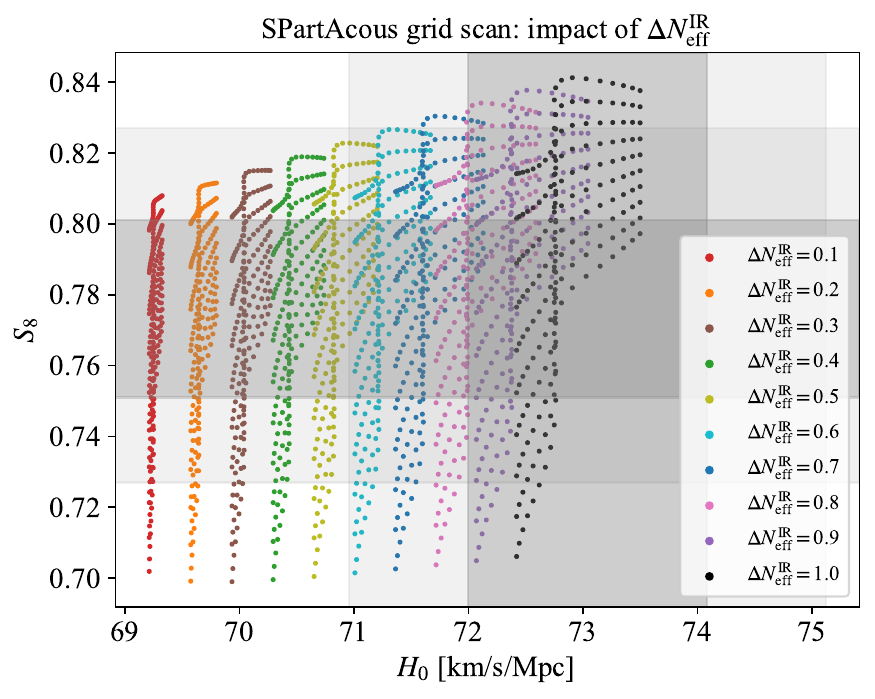}
	\includegraphics[width=.49\linewidth]{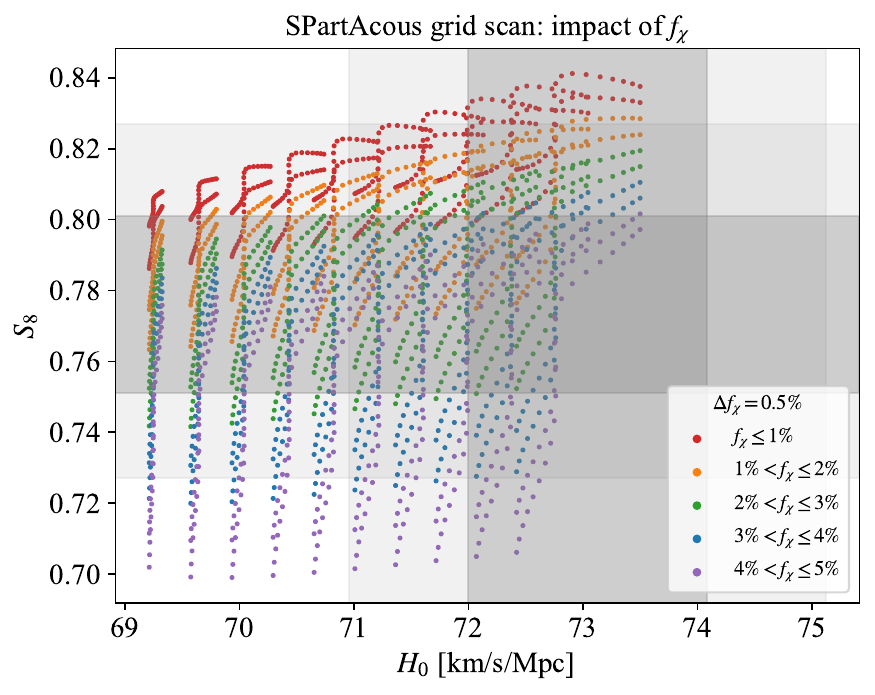}
	\includegraphics[width=.49\linewidth]{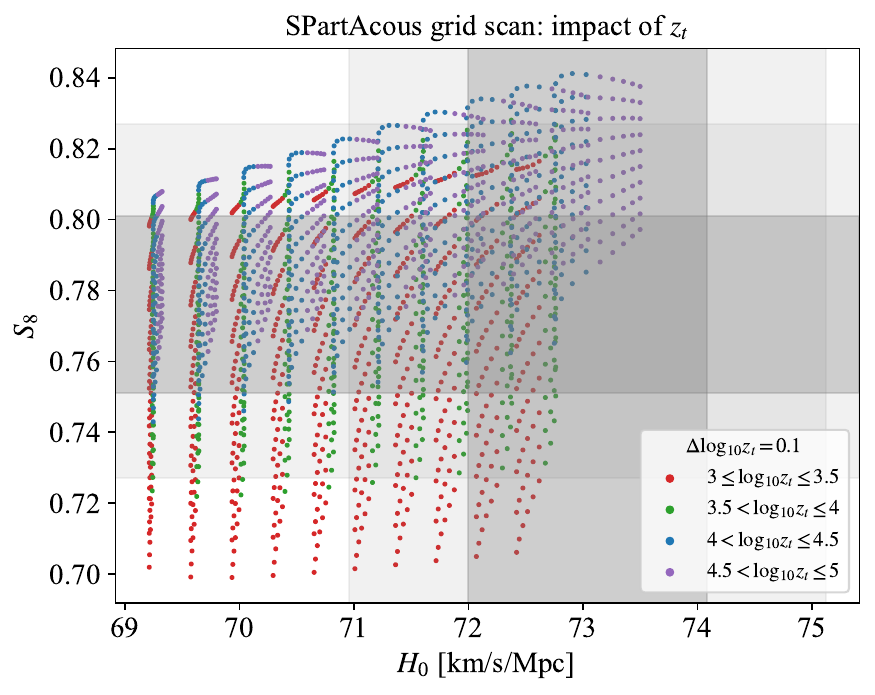}
	\includegraphics[width=.49\linewidth]{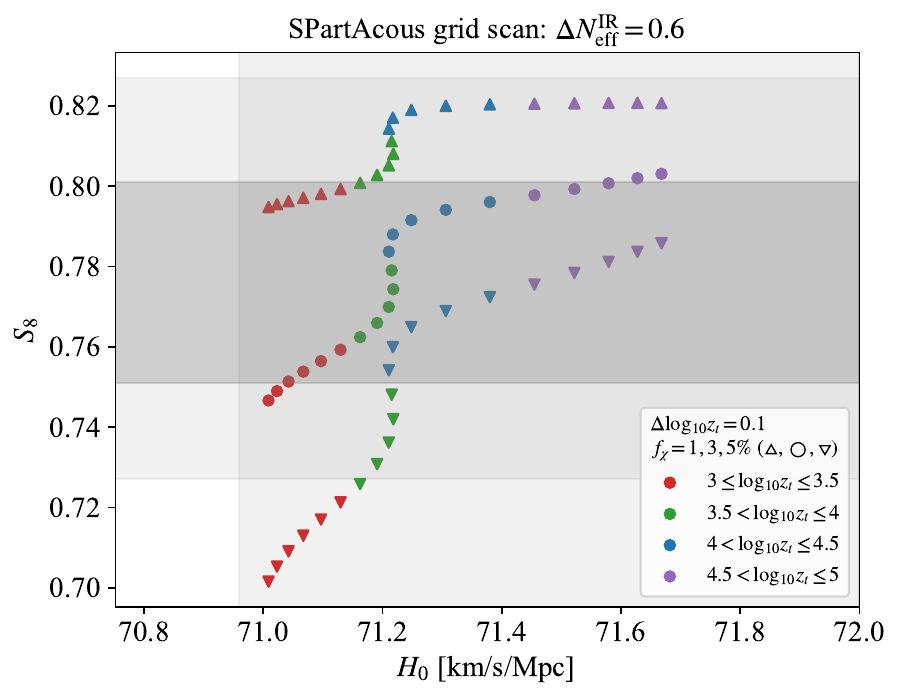}
	\caption{Scatter plots of the SPartAcous model's predictions for $\HO$ and $\Se$, compared to the $1\sigma$ and $2\sigma$ bands from the SH0ES ($\HO = 73.04 \pm 1.04~\HOunit$) \cite{Riess:2021jrx} and DES ($\Se = 0.775^{+0.026}_{-0.024}$) \cite{DES:2021wwk} collaborations, shaded in gray. The plotted points are obtained by varying the model parameters in the intervals $0.1 \leq \NIR \leq 1$, $0.5\% \leq f_\chi \leq 5\%$, and $3 \leq \log_{10} z_t \leq 5$, while fixing all the other cosmological parameters to the best fit of the WZDR model to the $\mathcal{D}+$ dataset from Ref.~\cite{Aloni:2021eaq}. The panels are color-coded according to varying $\NIR$ ({\bf top-left}), $f_\chi$ ({\bf top-right}), and $\log_{10} z_t$ ({\bf bottom-left}. In the {\bf bottom-right} panel, we zoom further into the bottom-left panel, taking $\NIR = 0.6$ and $f_\chi = 1\%$, $3\%$, and $5\%$).}
	\label{fig:scatter}
\end{figure}

In \Fig{fig:scatter} we show the ranges of $\HO$ and $\Se$ that can be obtained by scanning over this parameter region. In grey we show the $1\sigma$ and $2\sigma$ bands from the SH0ES ($\HO = 73.04 \pm 1.04~\HOunit$~\cite{Riess:2021jrx}) and DES ($\Se = 0.775^{+0.026}_{-0.024}$ \cite{DES:2021wwk}) collaborations. The point markers have been color-coded according to the values of the corresponding parameters across their ranges: the top-left panel colored based on $\NIR$, the top-right panel based on $f_\chi$, and the bottom-left based on $z_t$. From the top two panels in this figure, it is clear that by adjusting $\NIR$ and $f_\chi$ one can achieve larger values of $\HO$ while at the same time lowering $\Se$. In this regard the model behaves similarly to PAcDM. The impact of $z_t$ is more subtle, and can be better appreciated by studying the bottom-right panel. In it, we show the values of $\HO$ and $\Se$ for fixed $\NIR = 0.6$ and only three values for $f_\chi$ ($1\%$, $3\%$, and $5\%$) as $z_t$ varies over its range. As $z_t$ increases, the DR step takes place at earlier times such that SPartAcous becomes increasingly similar to just DR with $\DNeff = \NIR$ and without a step, except for very small scales that enter the horizon before $z_t$. Along this direction $\Se$ grows, as the scales that entered the horizon before $z_t$ experience a smaller period of DAO. Furthermore, the effect of the step on the comoving sound horizon disappears, and instead of effectively having a DR scenario with both a low $\NUV$ and a large $\NIR$ there is only a single value of $\DNeff = \NIR$. Since $\DNeff$ is positively correlated with $\HO$, one ends up with a larger $\HO$. Note that the worsening of $\Se$ as $z_t$ increases in order to improve $\HO$ can be compensated by having a larger $f_\chi$. In summary, SPartAcous can produce values of $\Se$ and $\HO$ well within the current experimental errors for the direct determination of these parameters.

\begin{figure}[tb]
	\centering
	\includegraphics[width=.49\linewidth]{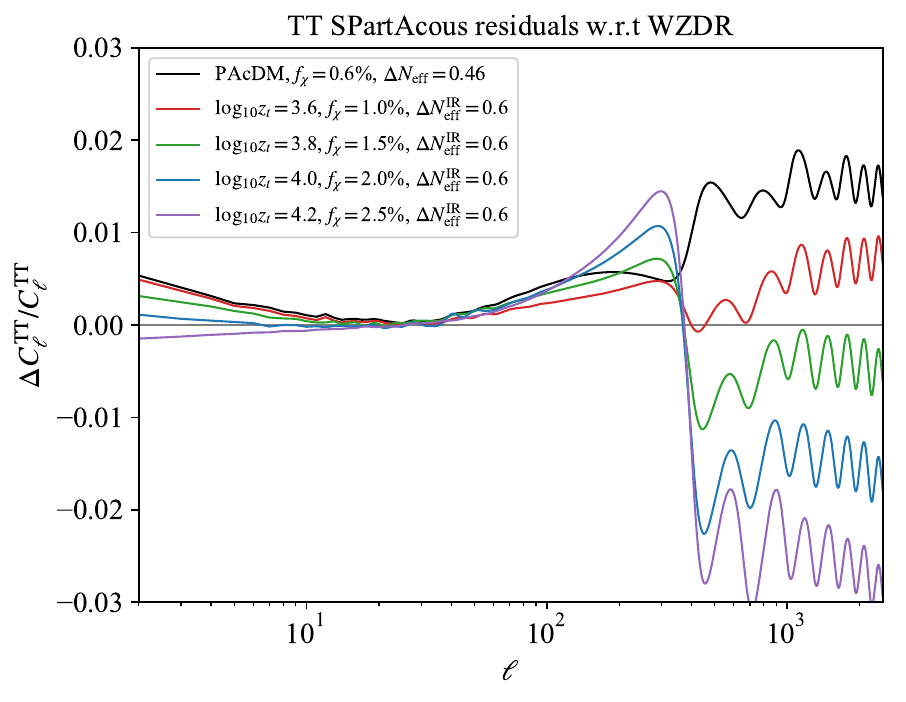}
	\includegraphics[width=.49\linewidth]{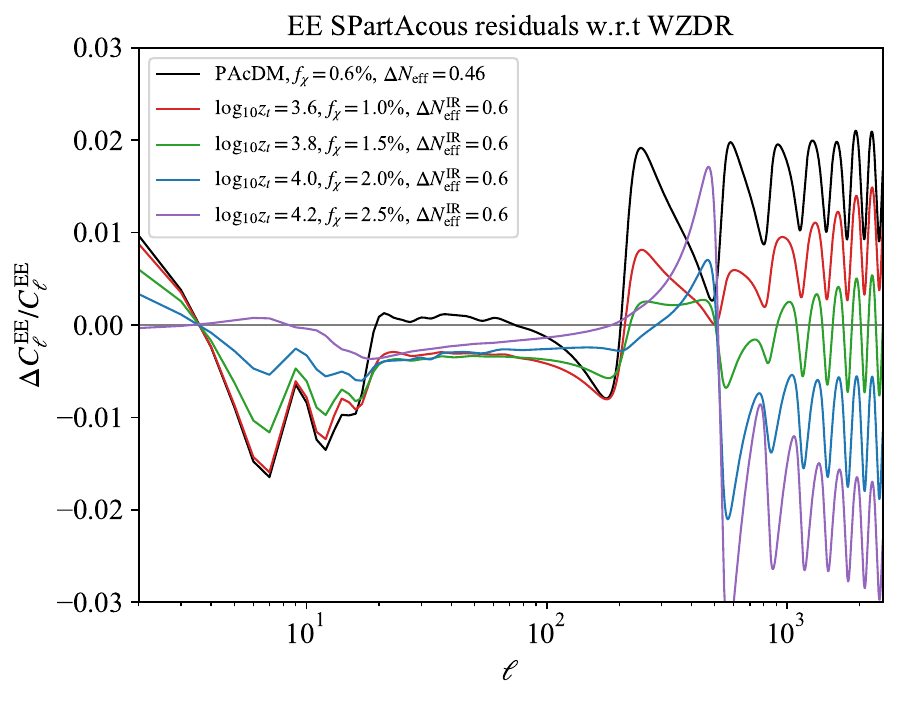}
	\caption{Residuals for the TT ({\bf left}) and EE ({\bf right}) CMB power spectra of the SPartAcous model compared to the best fit point of WZDR to the $\mathcal{D}+$ dataset in Ref.~\cite{Aloni:2021eaq}. The chosen values of the SPartAcous parameters $\{ \NIR, f_\chi, z_t \}$ are such that both $\HO$ and $\Se$ tensions are significantly reduced (more concretely $\HO > 71~\HOunit$ and $\Se$ within $1\sigma$ of DES). Note that for $\log_{10}(z_t) \sim 3.8$ the residuals are within less than $2\%$ of the best fit point of the WZDR model. For comparison, we also show in grey the residuals for the PAcDM-like limit of SPartAcous.}
	\label{fig:WZDR_residuals}
\end{figure}

\begin{figure}[tb]
	\centering
	\includegraphics[width=.49\linewidth]{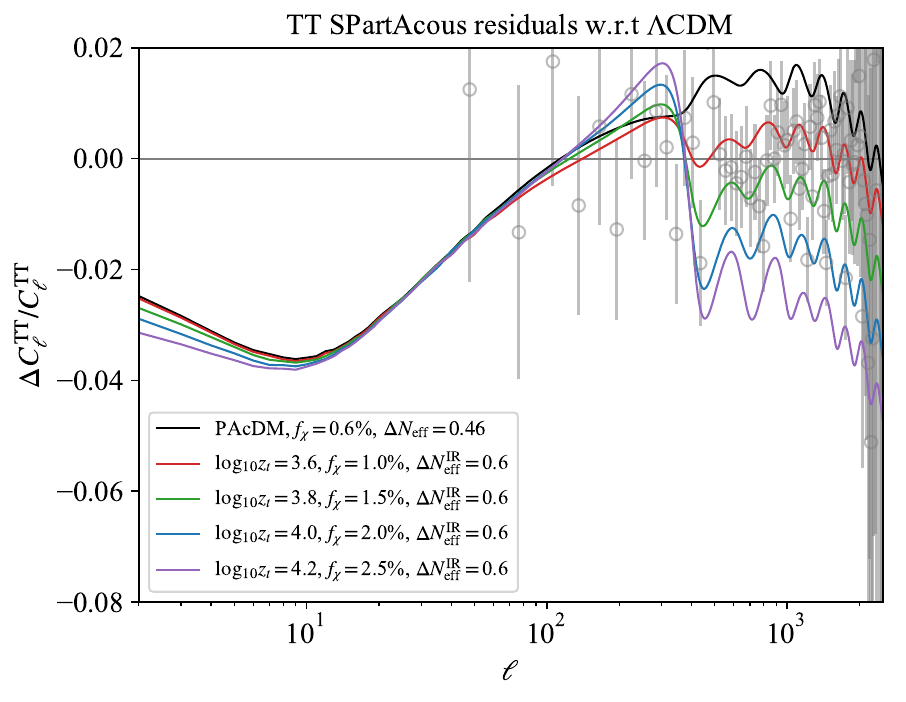}
	\includegraphics[width=.49\linewidth]{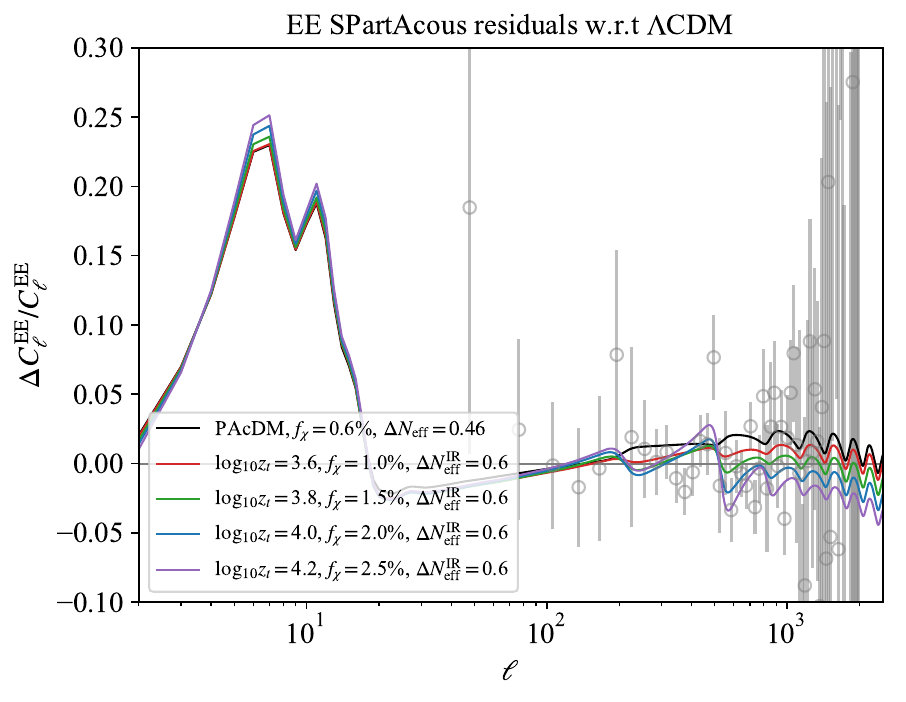}
	\caption{Residuals for the TT ({\bf left}) and EE ({\bf right}) CMB power spectra of the SPartAcous model compared to the best fit point of $\Lambda$CDM to the $\mathcal{D}+$ dataset in Ref.~\cite{Aloni:2021eaq}. The chosen values of the SPartAcous parameters $\{ \NIR, f_\chi, z_t \}$ are such that both $\HO$ and $\Se$ tensions are significantly reduced (more concretely $\HO > 71~\HOunit$ and $\Se$ within $1\sigma$ of DES). Note that for $\log_{10}(z_t) \sim 3.8$ the residuals are within $2\%$ of the best fit point of the WZDR model. For comparison, we also show in grey the residuals for the PAcDM-like limit of SPartAcous (obtained by taking a very late DR step, $z_t = 0.1$ in this figure).}
	\label{fig:LCDM_residuals}
\end{figure}

Having shown SPartAcous's promising potential to address the $\HO$ and $\Se$ tensions, it is of vital importance to investigate whether the model will also recover the good fit of the original WZDR model to CMB data. While we leave a full MCMC scan for upcoming work, we can get a sense of what these interactions do to the CMB from \Fig{fig:WZDR_residuals} and \Fig{fig:LCDM_residuals}. In these figures, we plot the TT and EE residuals from comparing SPartAcous to the WZDR best fit (Fig.~\ref{fig:WZDR_residuals}) and to the $\Lambda$CDM best fit (Fig.~\ref{fig:LCDM_residuals}), for a handful of $\{ \NIR, f_\chi, z_t \}$ parameters for which both the $\HO$ and $\Se$ tensions are significantly reduced (more concretely $\HO > 71~\HOunit$ and $\Se$ within $1\sigma$ of DES). We can see that the SPartAcous residuals with respect to WZDR are within $2 \%$, while the residuals with respect to the best fit of \lcdm~to Planck data (\Fig{fig:LCDM_residuals}) are also within $2\%$ over almost the entire $l$-range, except for small $l$'s which are much less constrained due to cosmic variance. Given the typical error bars for the CMB (see Fig.~\ref{fig:LCDM_residuals}), this is an encouraging sign that the model should still provide a very good fit to CMB data for transition redshifts in the neighborhood of $z_t \approx 10^{3.8}$.

After identifying a promising range of parameters based on \Fig{fig:scatter}, we use the MCMC sampler {\tt MontePython}~\cite{Brinckmann:2018cvx} to perform a simplified scan of the parameter space at a finer level, and compare the $\chi^2$ results to $\Lambda$CDM and WZDR. For this simplified scan, we fix all cosmological parameters to have the values used in the red curve of Figures~\ref{fig:WZDR_residuals} and~\ref{fig:LCDM_residuals}, except for $f_\chi$, which we use {\tt MontePython} to scan over. We use the following data for the {\tt MontePython} scan: the Planck 2018 TT, TE, EE and lensing likelihoods~\cite{Planck:2018vyg}, the BAO data from BOSS DR12~\cite{BOSS:2016wmc} and small $z$ observations~\cite{Beutler_2011, Ross:2014qpa}, the Pantheon Supernova data~\cite{Pan-STARRS1:2017jku}, the SH0ES measurement of $H_0$ from Ref.~\cite{Riess:2021jrx} and finally $S_8$ from KiDS-1000x~\cite{Heymans:2020gsg} and DES-Y3~\cite{DES:2021wwk}. The best fit point has $H_0 = 71.2~\HOunit$ and $S_8 = 0.806$, and yields a total $\chi^2 = 3834.99$, which improves upon the $\Lambda$CDM fit value of $\chi^2 = 3843.54$ (Table VI of Ref.~\cite{Joseph:2022jsf}) by $\Delta \chi^2 = -8.55$. In addition, in order to see whether the model can ease the tensions without including the $H_0$ and $S_8$ measurements themselves, we perform a second scan using only the CMB and BAO data, fixing the model parameters to be the same as in the WZDR best fit to their $\mathcal{D}$ dataset in Ref.~\cite{Aloni:2021eaq} (including $\NIR = 0.23$ and $z_t = 10^{4.3}$). We find that the best fit point in this second scan results in $\chi^2 = 3808.09$, comparable to WZDR ($\chi^2 = 3807.3$) and to $\Lambda$CDM, ($\chi^2 = 3808.5$ - Table VII of Ref.~\cite{Aloni:2021eaq}). The Hubble parameter at the best fit point of this second scan is $H_0 = 69.0~\HOunit$, which is a larger value than that of $\Lambda$CDM ($H_0 = 67.6~\HOunit$) and comparable to that of WZDR ($H_0 = 69.1~\HOunit$), and the SH0ES tension is eased by $\Delta \chi^2_{\rm SH0ES} = -8.12$. Thus, the SPartAcous model can accommodate larger values of $H_0$ without the inclusion of the SH0ES dataset. These results, obtained from a partial MCMC scan (over one parameter), show the promising nature of SPartAcous. A full scan will be performed in an upcoming paper~\cite{Buen-Abad:2023b}.

\begin{figure}[tb]
	\centering
	\includegraphics[width=.49\linewidth]{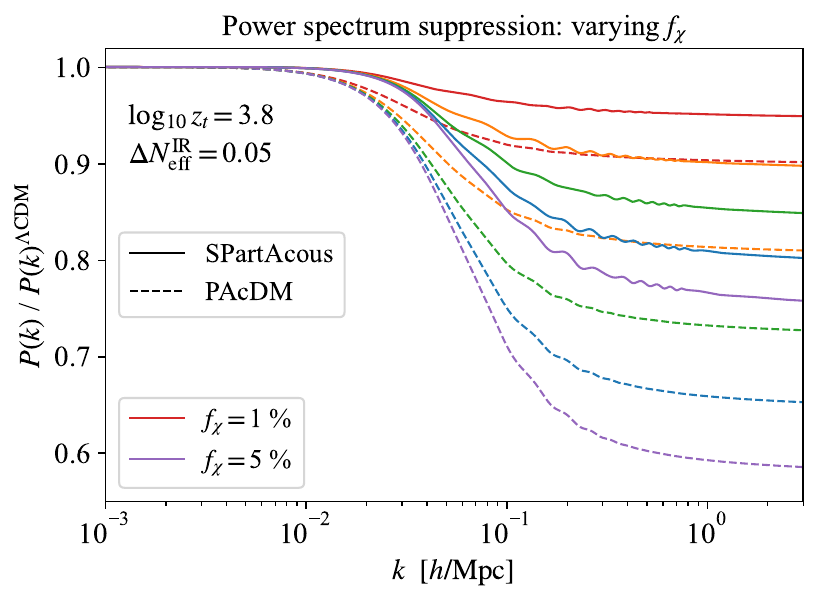}
	\includegraphics[width=.49\linewidth]{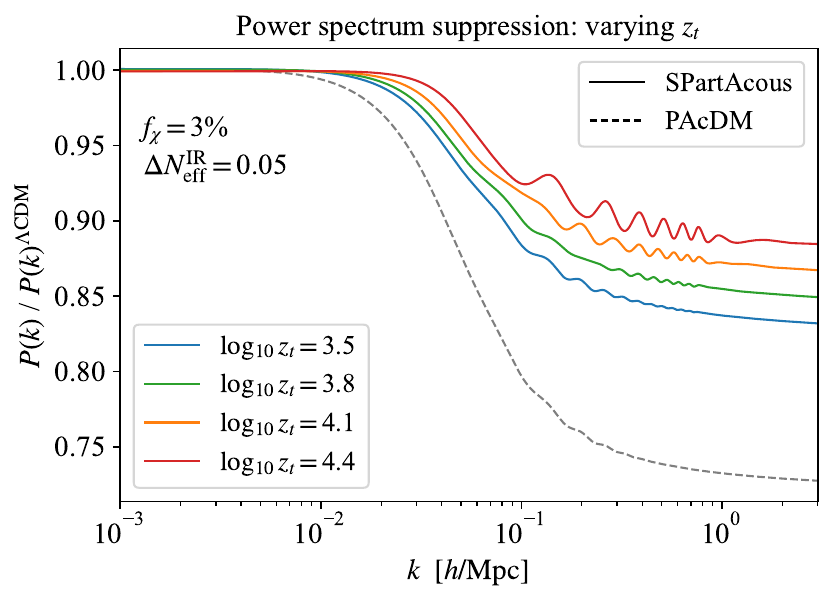}
	\caption{Ratio of the matter power spectrum of the SPartAcous model to that of \lcdm, for varying iDM fraction $f_\chi$ and fixed $z_t$ ({\bf left}), and varying $z_t$ with fixed $f_\chi$ ({\bf right}). We have fixed $\NIR = 0.05$, as well as the redshift $z_\eq$ of matter--radiation equality and the size of the CMB angular scales $\theta_s$ to their \lcdm~best fit values to Planck 2018 data \cite{Planck:2018vyg}, in order to better isolate the impact of $f_\chi$ and $z_t$. All other parameters have been fixed to their \lcdm~best fit values as well. For comparison we also show the matter power spectrum suppression found in PAcDM \cite{Chacko:2016kgg,Buen-Abad:2017gxg} for the same values of $\DNeff$ and $f_\chi$ (dashed curve(s) in both panels; recall that PAcDM has no $z_t$ parameter).}
	\label{fig:powerspectrum}
\end{figure}

In addition to providing a promising solution to both cosmological tensions, the SPartAcous model gives a qualitatively new template for the matter power spectrum, which can potentially be distinguished from those of other interacting dark sector scenarios \cite{Buen-Abad:2015ova, Lesgourgues:2015wza, Chacko:2016kgg, Buen-Abad:2017gxg} with future large scale structure measurements. Indeed, in \Fig{fig:powerspectrum} we show the effect of varying $f_\chi$ and $z_t$ on the linear matter power spectrum expressed as the ratio of the power spectrum in SPartAcous to that of $\Lambda$CDM. In order to isolate the impact of these parameters on the matter power spectrum, we have chosen a small amount of DR for illustrative purposes, $\NIR = 0.05$. One can then easily see that the effect of increasing $f_\chi$ is to decrease power at small scales, while $z_t$ controls where the transition to \lcdm-like behavior occurs. As $z_t$ increases and the DR step takes place at earlier times, the DAO also end earlier, a short time after $z_t$. As a result the stunted growth of matter overdensities is relegated to smaller scales.

This transition from interacting to collisionless DM in SPartAcous, dialed by $z_t$, is crucial in order to maintain a good fit to the CMB while attempting to address the $\HO$ and $\Se$ tensions. In PAcDM, increasing the amount of DR to values large enough to address the $\HO$ tension causes the decrease in the power spectrum to extend to larger wavelengths, well into the scales probed by the CMB and therefore highly constrained by data~\cite{Buen-Abad:2017gxg}. In SPartAcous, however, the shutting-off of the DR--DM interactions, associated with the step in the DR, decreases the impact that these interactions have on CMB observables. Effectively, the suppression in the matter power spectrum is limited to smaller scales, corresponding to those that entered the horizon before the step, leaving unaffected those scales most precisely measured by the CMB.

\begin{figure}[tb]
	\centering
	\includegraphics[width=.49\linewidth]{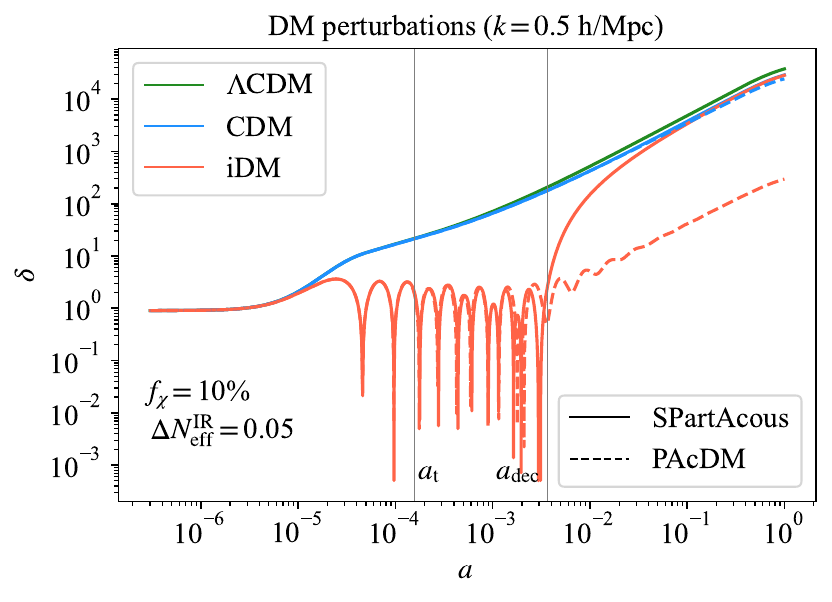}
	\includegraphics[width=.49\linewidth]{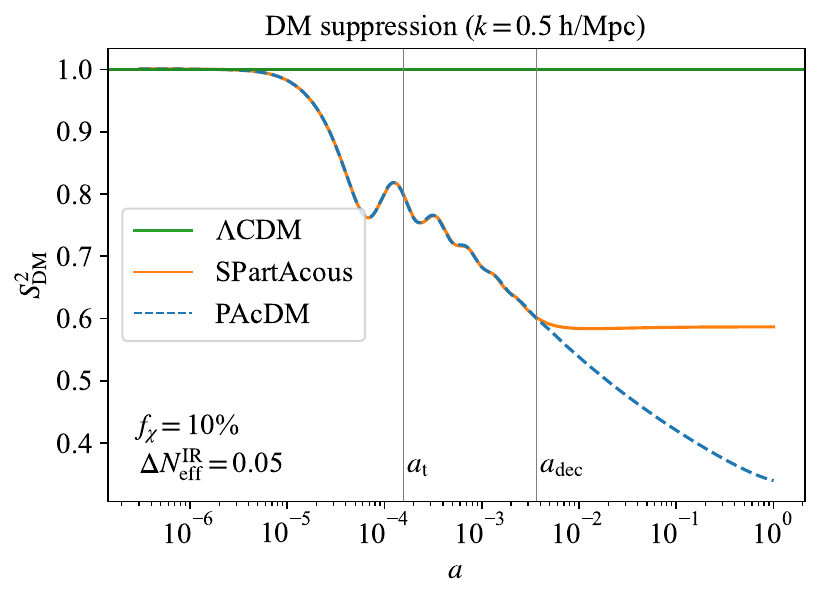}
	\caption{The evolution of a perturbation with $k=0.5~h/\Mpc$ is plotted as a function of $a$, comparing the PAcDM and SPartAcous models. We take $f_\chi = 10\%$, $\NIR = 0.05$ (for PAcDM, which has no DR step, this translates to $\DNeff = \NUV \approx 0.036$), and $\log_{10} \, z_t = 3.8$ for SPartAcous. We have taken a large $f_\chi$ value to enhance its effect for illustrative purposes. {\bf Left:} The evolution of $\delta$ for collisionless CDM, and for iDM in the two models is plotted. The acoustic oscillations of the $\delta_\idm$ perturbations, due to their coupling to the DR, are visible, starting as soon as $k$ enters the horizon. At a scale factor $a_t$, $\psi$ starts annihilating in SPartAcous, and at a scale factor $a_{\rm dec}$ (when $\Gamma = H$) iDM--DR decoupling occurs. Note that the amount of DR in the SPartAcous model changes around $a_t$, resulting in a relative phase shift between the iDM oscillations in SPartAcous and PAcDM. {\bf Right:} The $S_{\rm dm}^2$ suppression of the sum total of the DM perturbations (see \Eq{eq:suppr}) in SPartAcous and PAcDM is plotted relative to \lcdm.}
	\label{fig:perts}
\end{figure}

We can understand two noticeable features of this new template to the power spectrum, namely the origin of the DAO, and the smaller suppression of the power spectrum in SPartAcous compared to PAcDM for the same values of $f_\chi$ and $\NIR$, by studying the time-dependent behavior of the DM perturbations. In the left panel of \Fig{fig:perts}, we show the evolution of the $\delta_\cdm$ and $\delta_\idm$ perturbations for a single wavenumber, $k = 0.5~h/\Mpc$, for \lcdm, PAcDM, and SPartAcous. As can be seen for both PAcDM and SPartAcous, $\delta_\idm$ undergoes acoustic oscillations as soon as it enters the horizon, due to its tight coupling to the DR. This means that the iDM does not clump, and the gravitational potentials are shallower than in \lcdm, making $\delta_\cdm$ in SPartAcous and PAcDM grow at a slower rate than in \lcdm. At a redshift $a_{\rm dec}$, not far after the step at $a_t = (1+z_t)^{-1}$, the exponential factor in \Eq{eq:momentum-transfer} reduces $\Gamma / H$ below 1, and the $\chi$ particles decouple from the DR. After this, there is no DR pressure to sustain the iDM acoustic oscillations. Therefore, $\delta_\idm$ in SPartAcous starts growing due to the gravitational potential sourced by CDM, and catches up with $\delta_\cdm$. Once $\delta_\idm \approx \delta_\cdm$, both perturbations start growing as in a standard CDM scenario (and therefore faster than when iDM was undergoing acoustic oscillations) - this is analogous to when baryons decouple from photons shortly after recombination. While the decoupling takes places at the same time $a_{\rm dec}$ for all scales, perturbations with different $k$ scales will find themselves at different stages of their oscillations at this point in time (troughs, peaks or anything in between). As a result, the different $k$ modes will reach the $\delta_\idm \approx \delta_\cdm$ regime at different phases of their oscillation, and accordingly spend more or less time in the regime of faster growth. This imprints the DAO feature into the power spectrum as can be seen in Fig.~\ref{fig:powerspectrum}.

In contrast, for PAcDM, there is no mass threshold in the DR, and so no decoupling between the iDM and the DR takes place. Therefore, the pressure support from DR remains active and the $\delta_\idm$ continues oscillating. Although the perturbations in iDM eventually start to grow once the sound speed $c_s^2$ starts decreasing, which happens when the energy density in iDM exceeds that in DR, they only grow at the same rate as $\delta_\cdm$ and never catch up~\cite{Buen-Abad:2017gxg}. Consequently, they do not imprint their acoustic oscillations into the matter power spectrum. 

Note also that since the $\delta_\idm$ perturbations eventually catch up to $\delta_\cdm$, {\it both} kinds of DM contribute to the gravitational potential and consequently there is less power suppression in SPartAcous compared to PAcDM. From this point onward, the amount of clumping DM in SPartAcous is the same as that in \lcdm~ ($f_\chi$ of clumping iDM and $1-f_\chi$ of CDM), and both perturbations grow with the same rate as $\delta_\cdm$ in \lcdm. However, since there was a time prior to the step when this rate was not the same, there remains an overall suppression in the DM perturbations in SPartAcous with respect to \lcdm~ for short wavelengths, even if it is not as severe as in PAcDM. This can be understood in terms of the suppression in the DM perturbations, which can be parameterized in terms of \cite{Buen-Abad:2017gxg}
\beq\label{eq:suppr}
    S_{\rm dm}^2 \equiv \frac{\bl( f_\chi \delta_\idm + (1 - f_\chi) \delta_\cdm \br)^2\vert_{\rm model}}{\delta_\cdm^2\vert_{\lcdm}} \ .
\eeq
This has been plotted in the right panel of \Fig{fig:perts}, for both SPartAcous and PAcDM models. Shortly after $a_{\rm dec}$, $S_{\rm dm}^2$ for SPartAcous flattens out, which reflects the fact that $\delta_\cdm$ and $\delta_\idm$ now grow at the same rate as their \lcdm~counterparts. The fact that $S_{\rm dm}^2$ is larger in SPartAcous than in PAcDM is also responsible for the more moderate matter power spectrum suppression in the SPartAcous template compared to PAcDM.

It is clear from this discussion that the matter power spectrum of the SPartAcous model exhibits distinctive features for modes with $k$ in the range from $10^{-2} \, - \, 10^{-1}$ h/Mpc. With the expected improvements in LSS data and better modeling of the bias parameters, it may be possible to use this to distinguish SPartAcous from other scenarios that have been proposed to solve the $S_8$ tension. In addition, given the percent level sensitivity to $\DNeff$ and $\Delta N_\text{fluid}$ expected from CMB-S4, it should be sensitive to the $\mathcal{O}(1)$ step-like change in $\Delta N_\text{fluid}$ predicted by our model, which occurs during the CMB epoch. This would help distinguish this framework from other scenarios without such a mass threshold. Together, future CMB and LSS measurements may be able to distinguish SPartAcous from other solutions to the $H_0$ and $S_8$ tensions.

\section{Conclusions}
\label{sec:conclusions}

In this work we studied a new interacting dark sector model, SPartAcous (for {\it ``Stepped Partially Acoustic Dark Matter''}), which generalizes the Partially Acoustic Dark Matter (PAcDM) paradigm~\cite{Chacko:2016kgg} in a simple and fruitful way, via the introduction of a mass-threshold for a subcomponent of the DR. This threshold produces a step-like increase in $\Neff$, which was recently shown~\cite{Aloni:2021eaq} to substantially improve the $H_0$ tension. In addition, since the subcomponent of DR that interacts with DM is the one that becomes massive, this effectively turns off the interactions between DR and DM. This ensures that the main effect of the interactions is to create a partial suppression of the matter power spectrum at small scales, without affecting it at the larger scales for which there is greater CMB sensitivity. This suppression is precisely what the direct measurements of large-scale structure appear to prefer, allowing our model to address the $\Se$ tension.

We presented a simple realization of this scenario, identified a parameter point for which the evolution occurs as expected, and derived the relevant equations for the cosmological evolution of the new components. Using those results, we provided solid numerical evidence that this scenario captures the best features of both the PAcDM model and the original stepped DR model of~\cite{Aloni:2021eaq}, improving the $H_0$ tension while simultaneously addressing the $S_8$ tension. We also showed that this scenario provides a qualitatively new template for the matter power spectrum, with potentially observable new features compared to earlier interacting dark sector models that may serve as a smoking gun signature. This template exhibits a suppressed spectrum at small scales due to the large interactions between iDM and DR at early times, but a \lcdm-like spectrum at larger scales. In addition, evidence of these interactions is imprinted on the matter power spectrum in the shape of acoustic oscillations, in a way entirely analogous to the well-known baryon acoustic oscillations. As an added bonus, since at late times there are no further interactions between the DR and the DM, the impact of this template on non-linear scales can be conveniently studied with the same tools as \lcdm, such as N-body simulations and EFTofLSS, both of which are based on collisionless DM.

\hspace{1pc}

\noindent {\bf Note Added:} As this manuscript was being prepared two papers appeared~\cite{Schoneberg:2022grr,Joseph:2022jsf}, which combine the StepDR scenario with the very weak long-range type of interactions between DR and DM first discussed in Ref.~\cite{Buen-Abad:2015ova} to address the $\HO$ and $\Se$ problems. Both of these references deal with {\it weak} interactions between the DR and the {\it totality} of the DM, in the framework of the Wess-Zumino version of StepDR. In Ref.~\cite{Schoneberg:2022grr}, the interactions remain unchanged after the step, while in Ref.~\cite{Joseph:2022jsf}, the interaction rate starts to redshift as a steeper power of temperature after the step and quickly becomes irrelevant. Our paper differs significantly from both these works, since it is based on a framework in which DR has {\it strong} interactions with only a {\it subcomponent} of the total DM. Consequently, the change in the interaction rate after the step is exponentially sensitive to the temperature in our scenario, leading to a much sharper transition in matter power spectrum behavior. The resulting impact on the various cosmological observables, particularly the matter power spectrum, is therefore very different. In particular, the oscillatory feature arising from the dark acoustic oscillations is only present in our scenario. This may allow future LSS measurements to discriminate 
between these different scenarios.

\acknowledgments{
The authors thank Martin Schmaltz and Melissa Joseph for useful discussions. MBA thanks Stephanie Buen Abad for reviewing this manuscript. MBA, ZC and GMT are supported in part by the National Science Foundation under Grant Number PHY-1914731. ZC and GMT are also supported in part by the US-Israeli BSF Grant 2018236. The research of CK and TY is supported by the National Science Foundation Grant Number PHY-1914679.
}

\appendix

\section{Cosmological Evolution in the SPartAcous Model}
\label{sec:app}

We begin this appendix by summarizing the phase-space distribution (p.s.d.) function approach to describing the energy-momentum tensor of a cosmological fluid, and then move on to derive the relevant equations governing the evolution of the constituents of the SPartAcous model. Throughout this appendix we will follow closely the notation of Ref.~\cite{Buen-Abad:2017gxg}, which differs from that of the classic Ref.~\cite{Ma:1995ey} by a few factors of $2\pi$. We will be working in the conformal Newtonian gauge.

\subsection{Fluid Description from Phase Space}
\label{A1-fluid}

The p.s.d. function $f(\tau, \vx, \vp)$ for particles with $g$ degrees of freedom is normalized in a way such that their number density is given by,
\bea
    n(\tau, \vx) & = & g \int \! Dp ~ f(\tau, \vx, \vp) \ ,\\
    \text{with }\quad Dp & \equiv & \frac{\dd^3 p}{(2\pi)^3} \ .
\eea
Here $\tau$, $\vx$ are the time and space conformal coordinates, and $\vp$ the physical momentum of the particles in question. The Lorentz-invariant measure is defined as $\dd\Pi \equiv Dp/2E$, where $E = \sqrt{m^2 + p^2}$ is the energy of the particles.

In a universe that is, on average, homogeneous and isotropic with only small anisotropies, the p.s.d. of its constituent particles can be expressed as the sum of the background p.s.d. $f_0(\tau, p)$, which depends only on time and the magnitude of the particles' momenta, and perturbations about this background, $\Psi(\tau, \vx, \vp)$. Eliminating $\vx$ in favor of its Fourier conjugate $\vk$, and expressing all angular dependence in terms of $\mu \equiv \hat{p}\cdot\hat{k}$, we have
\beq\label{eq:psd_expand}
    f(\tau, \vec{p}, \vec{k}) = f_0(p, \tau) \bl( 1 + \Psi(\tau, p, k, \mu) \br) \ .
\eeq

It is convenient to define the comoving energy and momenta of the particles, given by $q = ap$ and $\epsilon = a E = \sqrt{q^2 + m^2 a^2}$ respectively, where $a(\tau)$ is the scale factor of the expansion of the Universe. In terms of these variables, the average number density, energy density, and pressure, are given by \cite{Ma:1995ey},
\bea
    \overline{n} & \equiv & a^{-3} \frac{g}{2\pi^2} \int \dd q ~ q^2 f_0 \ , \label{eq:n}\\
    \overline{\rho} & \equiv & a^{-4} \frac{g}{2\pi^2} \int \dd q ~ q^2 f_0 ~ \epsilon \ , \label{eq:rho}\\
    \overline{P} & \equiv & a^{-4} \frac{g}{2\pi^2} \int \dd q ~ q^2 f_0 ~ \frac{q^2}{3\epsilon} \label{eq:p}\ ,
\eea
where for the sake of brevity we have omitted the arguments of all the functions. From now on we drop the overline; whether our expressions refer to the average quantities or the total ones (including perturbations) should be clear from the context.

Small inhomogeneities of these macroscopic quantities can be derived from the perturbations $\Psi$ about the background p.s.d.. It is more expedient to decompose these perturbations in terms of their Legendre coefficients $\Psi_l$ based on the angular variable $\mu$. Indeed, with the aid of the Legendre polynomials $P_l(\mu)$, we have \cite{Ma:1995ey},
\bea
    \Psi(q, k, \mu, \tau) & \equiv & \sum\limits_{l=0}^\infty (-i)^l (2l+1) P_l(\mu) \Psi_l(q, k, \tau) \\
    \delta \rho \equiv \rho \, \delta & \equiv & a^{-4} \frac{g}{2\pi^2} \int \dd q ~ q^2 f_0 \Psi_0 ~ \epsilon \ , \label{eq:delta} \\
    \delta P & \equiv & a^{-4} \frac{g}{2\pi^2} \int \dd q ~ q^2 f_0 \Psi_0 ~ \frac{q^2}{3\epsilon} \ , \label{eq:deltaP} \\
    u \equiv (\rho+P)\theta & \equiv & a^{-4} k \frac{g}{2\pi^2} \int \dd q ~ q^2 f_0 \Psi_1 ~ q \ . \label{eq:theta}
\eea
In general there are an infinite number of $l$-modes $\Psi_l$ to keep track of.

The cosmological evolution of a certain energy constituent of the universe is determined by the dynamics of its p.s.d.. The equation governing this dynamics is the Boltzmann equation, which can be schematically written as,
\beq
    \hat{L}[f] = \hat{C}[f] \ ,
\eeq
where $L$ denotes the {\it Liouville operator}, which describes the free streaming (along geodesics) of the particles in question, and $\hat{C}$ the {\it collision operator}, which accounts for the impact that any interactions may have on the p.s.d. of the particles.

The metric in conformal Newtonian gauge is given by
\begin{equation}
    ds^2 = a^2(\tau)\left(-(1 + 2 \psi)d\tau^2 + (1- 2 \phi)d\vec x^2\right) \, .
\end{equation}
In this gauge the Boltzmann equation is given by,
\beq\label{eq:boltz_eq}
    \dot{f} + i k \mu \frac{q}{\epsilon} f + \frac{\partial f}{\partial \ln q} \bl( \dot{\phi} - i k \mu \frac{\epsilon}{q} \br) = \frac{a^2}{\epsilon}(1 + \psi) ~ C(\tau, p, k, \mu) \ ,
\eeq
where the dot denotes derivatives with respect to $\tau$ and $C$ stands for the collision term.\footnote{Whether $\psi$ denotes the time-like metric perturbations in the Newtonian gauge or the massive DR constituent in SPartAcous should be clear from context.} The specific form of $C$ is model-dependent, and involves the square of the scattering amplitude for collisions of the particles in question with any other species with which they interact. For the case of 2-to-2 scattering with momenta $p_a p_b \rightarrow p^\prime_a p^\prime_b$, the collision term for the p.s.d. of species $a$ reads,
\beq
    C_{a} = \frac{g_a}{2} \int\! \dd\Pi_b \dd\Pi_a^\prime \dd\Pi_b^\prime ~ \overline{\vert \mathcal{M}_{ab\rightarrow a^\prime b^\prime} \vert^2} ~ (2\pi)^4 \delta^2(p_a + p_b - p_a^\prime - p_b^\prime) ~ F(p_a, p_b, p^\prime_a, p^\prime_b) \ ,
\eeq
where $g_i$ stands for the degrees of freedom in species $i$,
\beq
    \overline{\vert \mathcal{M}_{ab\rightarrow a^\prime b^\prime} \vert^2} \equiv \frac{1}{g_a g_b} \sum\limits_{\rm states} \vert \mathcal{M}_{ab\rightarrow a^\prime b^\prime} \vert^2
\eeq
is the total scattering amplitude squared, averaged over initial states, and
\beq
 F(p_a, p_b, p^\prime_a, p^\prime_b) = f(p_a^\prime)f(p_b^\prime)(1 \pm f(p_a))(1 \pm f(p_b)) - (p \leftrightarrow p^\prime)
\eeq
accounts for the phase space distributions and quantum statistics of the $a$ and $b$ particles taking part in the scattering.

\subsection{Simplifying Assumptions}
\label{A2-simple}

From \Eq{eq:boltz_eq}, one could in principle determine the evolution of macroscopic quantities such as $\rho$, $P$, $\delta$, and $\theta$. To achieve this, one can multiply this equation by the various integrands used in \Eqst{eq:rho}{eq:p} and \Eqst{eq:delta}{eq:theta} to define these macroscopic quantities, and then integrate with respect to $q$. However, there are in general two challenges in this approach: the infinity of $l$-modes to keep track of, and the fact that the resulting integrals over comoving momenta $q$ in \Eq{eq:boltz_eq} may not reduce, in each of their terms, to the closed forms in \Eqst{eq:n}{eq:theta}.

Nevertheless, the expressions simplify in some limiting cases. For example, for highly non-relativistic particles, which have p.s.d. with typical momentum much smaller than the mass ($p \ll m$), one needs to keep track of only the modes $l=0$ and $l=1$, the higher ones being suppressed \cite{Ma:1995ey}. This will be the case for the iDM component in our SPartAcous model. 

There is another important case where simplifications take place, and that is for a fluid with sufficiently large self-interactions so as to keep itself in local thermal equilibrium. In this case, any small anisotropies can be thought of as tiny deviations from the thermal equilibrium background distribution $f_0$,
\beq\label{eq:psd_thermal}
    f_0(\tau, p) = \frac{1}{e^{E(p, \tau)/T(\tau)} \pm 1} \ ,
\eeq
where the $+/-$ signs correspond to Fermi-Dirac and Bose-Einstein distributions respectively. Under this {\it ``local thermal equilibrium''} approximation \cite{Cyr-Racine:2015ihg, Buen-Abad:2017gxg}, the $\Psi$ perturbations about the thermal distribution $f_0$ can be expressed in terms of local variations of the temperature and bulk-velocity of the fluid, $\delta T(\tau, \vx) = T(\tau) ~ \alpha(\tau, \vx)$ and $\vec{v}(\tau, \vx) = \vec{\nabla} \beta(\vx)$.\footnote{The Helmholtz decomposition theorem and the fact that we are only interested in scalar perturbations (and therefore only in the irrotational components of the velocity) mean that we can always express $\vec{v}$ as the gradient of a scalar perturbation function $\beta$.} Crucially, there is no dependence on the direction $\hat{q}$ of the particles. Therefore, upon going to Fourier space, we have
\beq
\label{eq:anisotropies}
    f = \frac{1}{e^{\frac{E - \vec{p}\cdot\vec{v}}{T(1 + \alpha)}} \pm 1} \approx f_0 \bl[ 1 - \bl( \alpha + i k \mu ~ \beta ~ \frac{q}{\epsilon} \br) \frac{\partial \ln f_0}{\partial \ln \epsilon} \br] \ .
\eeq
From this it is clear that
\bea\label{eq:psd_perts2}
    \Psi & = & - \bl(\alpha + i k \mu \beta \frac{q}{\epsilon} \br) \frac{\partial \ln f_0}{\partial \ln \epsilon} \ , \quad \text{and therefore} \label{eq:Psil} \\
    \Psi_0 & = & - \alpha \frac{\partial \ln f_0}{\partial \ln \epsilon} \ , \label{eq:Psi0}\\
    \Psi_1 & = & \frac{k \beta}{3} \frac{q}{\epsilon} \frac{\partial \ln f_0}{\partial \ln \epsilon} \ , \label{eq:Psi1}\\
    \Psi_{l \geq 2} & = & 0 \ .\label{eq:Psi2+}
\eea
We can then see that the assumption of local thermal equilibrium (meaning that deviations from the average thermal distribution can be written in terms of local, $\vec{q}$-independent perturbations about the thermal distribution) is equivalent to assuming that the fluid in question has large self-interactions, which kill off the higher moments $\Psi_l$ for $l \geq 2$ \cite{Cyr-Racine:2015ihg}. Note that this is the case for the DR in the SPartAcous model.

Substituting \Eqs{eq:Psi0}{eq:Psi1} into \Eqs{eq:delta}{eq:theta} we arrive at the following very useful expressions,
\bea
    \alpha = \delta/4 \ , &\quad& \beta  = - k^{-2} \theta \ , \label{eq:alpha_beta}\\
    \Rightarrow \quad \Psi_0 & = & -\frac{\delta}{4} \frac{\partial \ln f_0}{\partial \ln \epsilon} \ , \label{eq:newPsi0}\\
    \Psi_1 & = & -\frac{\theta}{3 k} ~ \frac{q}{\epsilon} \frac{\partial \ln f_0}{\partial \ln \epsilon} \ .\label{eq:newPsi1}
\eea
These are valid for fluids in local thermal equilibrium.

\subsection{SPartAcous: Background Equations}
\label{A3-bg}

In the SPartAcous model we have three sub-components: a heavy charged scalar $\chi$ (which makes up the iDM), a massless gauge boson $A$, and a light charged fermion $\psi$. Each of these species has its own p.s.d.. The fact that the $A$ and $\psi$ particles have large interactions among themselves (see \Sec{sec:cosmo}) means that they efficiently exchange energy. As a result, they share the same temperature $T_d$, and their corresponding background thermal distributions are then given by
\beq\label{eq:dr_psd}
    f_{A0}(q) = \frac{1}{1 - e^{q/(aT_d)}} \ , \qquad f_{\psi0}(q) = \frac{1}{1 + e^{\epsilon/(aT_d)}} \ .
\eeq
In going from here to a macroscopic fluid description of the DR, we can simply consider the tightly-coupled $A$--$\psi$ system as a single DR fluid in local thermal equilibrium with itself, as described at the end of the previous section. As such, macroscopic quantities such as $\rho_\dr$ or $\delta P_\dr$ are simply given by the sum of the corresponding quantities for the individual $A$ and $\psi$ systems, which follows from \Eq{eq:dr_psd} applied to \Eqst{eq:rho}{eq:theta}. For example, $u_\dr = u_A + u_\psi$.

With this in mind, we are now ready to determine $\rho_\dr$ and $P_\dr$ as a function of the DR temperature $T_d$. We can obtain closed analytic formulas if we further assume a Boltzmann distribution $f_{0} = e^{-E/T_d}$ for both $A$ and $\psi$, following the authors of Ref.~\cite{Aloni:2021eaq}. This is a very good approximation, valid to better than 8\%, and we follow it for the rest of this paper. For the reader's convenience we summarize their results below, with the number of degrees of freedom appropriate to our model:
\bea
    \rho_\dr & = & g_*^{\rm IR} \rho_B (T_d) \bl( 1 + r_g \hat{\rho}(x) \br) \ , \label{eq:rho_dr} \\
    P_\dr & = & \frac{1}{3} g_*^{\rm IR} \rho_B (T_d) \bl( 1 + r_g \hat{p}(x) \br) \ , \label{eq:P_dr} \\
    w \equiv \frac{P_\dr}{\rho_\dr} & = & \frac{1}{3}\frac{1 + r_g \hat{p}(x)}{1 + r_g \hat{\rho}(x)} \ . \label{eq:w_dr}
\eea
Here $r_g \equiv (g_*^{\rm UV} - g_*^{\rm IR})/g_*^{\rm IR} = 7/4$, where $g_*^{\rm UV} = 11/2$ is the effective number of degrees of freedom in DR at high temperatures and $g_*^{\rm IR} = 2$ is the number of degrees of freedom left in the DR at low temperatures, after the dark fermions have annihilated away, $\rho_B(T_d) = \frac{\pi^2}{30} T_d^4$ is the energy density of a single bosonic degree of freedom, $x \equiv m_\psi / T_d$, and
\bea
    \hat{\rho}(x) & \equiv & \frac{x^2}{2}K_2(x) + \frac{x^3}{6}K_1(x) \ , \label{eq:rho_hat} \\
    \hat{p}(x) & \equiv & \frac{x^2}{2}K_2(x) \ , \label{eq:p_hat}
\eea
with $K_n(x)$ the $n$-th order modified Bessel function of the second kind.

The evolution of the DR temperature $T_d = m_\psi/x$ can be obtained by solving for $x(a)$ from the following equation,
\beq
    \bl( \frac{x a_t}{a} \br)^3 = 1 + \frac{r_g}{4}\bl( 3 \hat{\rho}(x) + \hat{p}(x) \br) \ , \label{eq:entropy}
\eeq
which follows from the conservation of entropy in the DR. Here $a_t \equiv \frac{1}{1+z_t} \equiv T_{d0}/m_\psi$ is the scale factor at which the step begins and the $\psi$ particles start to become non-relativistic.

Finally, the energy density in dark radiation, \DNeff, can be written as:
\beq\label{eq:DNeff}
    \DNeff(x) \equiv \frac{\rho_\dr}{\rho_{1\nu}} = \NIR \frac{1 + r_g \hat{\rho}(x)}{\bl( 1 + r_g \bl( \frac{3}{4} \hat{\rho}(x) + \frac{1}{3} \hat{p}(x) \br) \br)^{4/3}} \ ,
\eeq
where $\NIR = \frac{\rho_A}{\rho_{1\nu}}\bl.\br\vert_{T_d \ll m_\psi} = \frac{2}{\frac{7}{4} \left( \frac{4}{11} \right)^{4/3}} \, \bl( \frac{T_{d0}}{T_0} \br)^4 \approx 4.4 \bl( T_{d0}/T_0 \br)^4$, and $\NUV = \NIR/(1 + r_g)^{1/3}$. Note the complicated evolution as a function of redshift, as the DR temperature drops below the mass threshold $m_\psi$.

Finally, we need the sound speed of the DR, which for adiabatic perturbations is given by $c_s^2 \equiv \delta P_\dr/\delta \rho_\dr = P^\prime(x)/\rho^\prime(x)$:
\beq
    c_s^2(x) = \frac{1}{3}\frac{1 + r_g \bl( \hat{p}(x) - \frac{x}{4} \hat{p}^\prime(x) \br)}{1 + r_g \bl( \hat{\rho}(x) - \frac{x}{4} \hat{\rho}^\prime(x) \br)} \ . \label{eq:cs2}
\eeq
For more details, see Appendix A of Ref.~\cite{Aloni:2021eaq}.

The temperature of iDM evolves with time as \cite{Cyr-Racine:2015ihg, Bringmann:2006mu},
\beq
\frac{\dd T_\chi}{\dd\tau} + 2 \mathcal H T_\chi - \dot Q_{\psi \chi} (T_\mathrm{dr} - T_\chi) = 0 \, ,
\eeq
where $\dot Q_{\psi \chi}$ refers to the heating rate between the iDM and DR, given in \Eq{eq:Qchipsi}. In a manner analogous to the case of baryons and photons, our choice of parameters guarantees that $\chi$ is thermally tightly coupled to $\psi$, while $\psi$ and $A$ exchange heat efficiently far below $m_\psi$. We therefore have $(T_\mathrm{dr} - T_\chi) / T_\mathrm{dr} \ll 1$ until $\psi$ is frozen out.

\subsection{SPartAcous: Perturbation Equations}
\label{A4-perts}

The equations governing the evolution of the cosmological perturbations $\delta$ and $\theta$ can be derived from \Eq{eq:boltz_eq} by expanding the p.s.d. to first order in the perturbations $\Psi$. Decomposing the result into the $l$-modes $\Psi_l$ yields the equations known as the {\it ``Boltzmann hierarchy''} or the {\it ``Boltzmann ladder''},
\bea\label{eq:boltz_hier_1}
    && f_0 \bl( \dot{\Psi}_l + k \frac{q}{\epsilon} \bl( \frac{l+1}{2l+1}\Psi_{l+1} - \frac{l}{2l+1}\Psi_{l-1} \br) + \frac{\partial \ln f_0}{\partial \ln q} \bl( \dot{\phi} \delta_{l0} + \frac{k}{3}\frac{\epsilon}{q} \psi \delta_{l1} \br)  \br) \nn\\
    && \qquad = \frac{a^2}{2 \epsilon} (-i)^{-l} \int\limits_{-1}^{1} \dd \mu P_l(\mu) \ C^{(1)}(q) \ ,
\eea
where $C^{(1)}(q)$ is the collision operator to first order in the p.s.d. perturbations.

For the light $\psi$ particles in the DR, which have interactions with $\chi$, we can further simplify the Boltzmann hierarchy (working in the $m_\chi \gg m_\psi, \, p_\psi$ limit and following closely the appendices of Ref.~\cite{Cyr-Racine:2015ihg}),
\bea\label{eq:boltz_hier_2}
    && f_{\psi\,0} \bl( \dot{\Psi}_{\psi\, l} + k \frac{q}{\epsilon} \bl( \frac{l+1}{2l+1}\Psi_{\psi \, l+1} - \frac{l}{2l+1}\Psi_{\psi\, l-1} \br) + \frac{\partial \ln f_{\psi\, 0}}{\partial \ln q} \bl( \dot{\phi} \delta_{l0} + \frac{k}{3}\frac{\epsilon}{q} \psi \delta_{l1} \br)  \br) \nn\\
    && \qquad = - \rho_\chi \, \frac{a}{16\pi m_\chi^3} f_{\psi\,0} \bl( \frac{q^2}{\epsilon^2} \frac{\partial \ln f_{\psi\, 0}}{\partial \ln \epsilon} \, \delta_{l1} \, \Delta_l \, \frac{\theta_\chi}{3k} + \frac{q}{\epsilon} \, \Delta_l \, \Psi_{\psi\, l} \br) - \Lambda_{A\psi\,l} \ ,
\eea
where $\Lambda_{A\psi\,l}$ stands for the complicated collision term from $A$--$\psi$ scattering, and $\Delta_l$ is the transfer scattering amplitude squared for $\chi$--$\psi$ scattering,
\bea
    \Delta_l & \equiv & \frac{1}{2} \int\limits_{-1}^{1} \dd\tilde{\mu} ~ (1 - P_l(\tilde{\mu})) \overline{\bl\vert \mathcal{M}_{\chi\psi} \br\vert^2}_* \ , \label{eq:Deltal} \\
    * & : & s=m_\chi^2 + 2m_\chi E; \ t = p^2(\tilde{\mu}-1) \ .
\eea
Here $\tilde{\mu}$ parameterizes the angle between incoming and outgoing $\psi$'s ($\tilde{\mu} \equiv \hat{p}_\psi \cdot \hat{p}_\psi^\prime$). Note that the $\chi$--$\psi$ collision term in \Eq{eq:boltz_hier_2} vanishes for $l=0$. With the aid of \Eqs{eq:newPsi0}{eq:newPsi1} and the definitions \Eqst{eq:delta}{eq:theta}, we can then find the equations for $\delta\dot\rho_\psi$ and $\dot u_\psi$,
\bea
    \delta\dot\rho_\psi & = & - 3 \mH \bl( \delta\rho_\psi + \delta P_\psi \br) - u_\psi + 3 \dot\phi \bl( \rho_\psi + P_\psi \br) + ... \ , \label{eq:dot_rho_psi} \\
    \dot u_\psi & = & - 4 \mH u_\psi + k^2 \delta P_\psi + k^2 \psi (\rho_\psi + P_\psi) + \rho_\chi a \Gamma \bl( \theta_\chi - \theta_\psi \br) + ... \ , \label{eq:dot_u_psi}\\
    \Gamma & \equiv & \frac{a^{-4} g_\psi}{24 \pi^3 m_\chi^3} \int \! \dd q \, q^4 \, \bl( -\frac{1}{4} \frac{\partial f_{\psi0}}{\partial q} \Delta_1 \br) \ . \label{eq:Gamma_1}
\eea
The equations for the $A$ fluid are analogous to these, except without the $\Gamma$ term (since the $A$--$\chi$ Compton scattering is suppressed by higher powers of $T_d/m_\chi$ and therefore much smaller, see \Sec{sec:model} and \Sec{sec:cosmo}). The ellipses on the right-hand side of these equations denote any possible contributions from the $A$--$\psi$ collision term $\Lambda_{A\psi}$. As we have been discussing throughout this work (see \Sec{sec:model}), we are interested in the region of parameter space where the coupling between the $A$ and $\psi$ components is very large. As a result, the $A$ and $\psi$ fluids are in local thermal equilibrium, which means that they share a common temperature $T_d$ as well as the local variations $\alpha$ and $\beta$ in temperature and bulk velocity respectively, which parameterize the small perturbations about their background p.s.d. $f_0$; see \Eq{eq:anisotropies}. In particular, we have
\bea
    \theta_\psi  & = & -k^2 \beta = \theta_A \\
    \Rightarrow u_\dr \equiv \bl(\rho_\dr + P_\dr \br) \theta_\dr  & \equiv & u_\psi + u_A \nonumber\\
    & = & \bl( \rho_\psi + P_\psi \br)\theta_\psi + \frac{4}{3} \rho_A \theta_A \nonumber\\
    & = & \bl( \rho_\dr + P_\dr \br)\theta_\psi \\
    \Rightarrow \quad \theta_\dr & \equiv & \theta_\psi \ .
\eea
Since $\dot u_\dr = \dot u_\psi + \dot u_A$ and the ellipses in \Eqs{eq:dot_rho_psi}{eq:dot_u_psi} have the opposite sign to those for the corresponding $A$ equations (because of energy-momentum conservation in the $\psi$--$A$ system), we can write  equations for the evolution of $\delta_\dr \equiv \delta\rho_\dr \rho_\dr$ and $\theta_\dr \equiv u_\dr/(\rho_\dr + P_\dr)$,
\bea
    \dot \delta_\dr & = &  - (1+w)(\theta_\dr - 3 \dot \phi) - 3 \mH \left(c_s^2 - w \right) \delta_\dr \, ,\\
	\dot \theta_\dr & = & - \left[ (1-3w) \mH  + \frac{\dot w}{1+w} \right] \theta_\dr + k^2 \left(\frac{c_s^2}{1+w}  \delta_\dr +\psi \right) \nonumber\\
	& & + \frac{\rho_\idm}{\rho_\dr (1+w)} \, a \Gamma (\theta_\idm - \theta_\dr) \ ,
\eea
where $w$ and $c_s^2$ are the equation of state and the sound speed of the DR, given by \Eqs{eq:w_dr}{eq:cs2}. Repeating the same exercise for the very massive $\chi$ particles, we obtain
\bea
    \dot \delta_\idm & = &  - \theta_\idm + 3 \dot \phi  \, , \\
    \dot \theta_\idm & = & - \mathcal{H} \theta_\idm + k^2 \psi + a \Gamma \left( \theta_\dr - \theta_\idm \right) \ .
\eea

\subsection{The Momentum-Exchange Rate}
\label{A5-rate}

All that is left to do is to compute the momentum-exchange rate $\Gamma$, defined in \Eq{eq:Gamma_1}. For the Lagrangian given in \Eq{eq:lagrangian}, the averaged amplitude squared for $\chi\psi \rightarrow \chi\psi$ scattering is given by
\bea\label{eq:ampl2}
    \overline{\bl\vert \mathcal{M}_{\chi\psi} \br\vert^2} & = & \frac{4 g_d^4 ((s - m_\chi^2 - m_\psi^2)^2 + t\bl( s - m_\psi^2 \br))}{t^2} \nonumber\\
    & = & 32 \pi^2 \alpha_d^2 m_\chi^2~ \frac{2 ( m_\psi^2 + p^2 ) + (\tilde{\mu} - 1) p^2}{(\tilde{\mu}-1)^2 p^4} \,
\eea
where $g_d$ is the gauge coupling of the dark $U(1)$ gauge bosons $A$, $\alpha_d \equiv g_d^2/(4\pi)$, and we have taken $s = m_\chi^2 + 2m_\chi E$ and $t = p^2(\tilde{\mu}-1)$ in the $m_\chi \gg m_\psi, \, p$ limit. Note that we have averaged over the processes involving particles and anti-particles for both $\psi$ and $\chi$.

We are now ready to compute $\Delta_1$ from \Eq{eq:Deltal}. In order to do this, we need to regularize the $\tilde{\mu}$ integral, since this amplitude diverges for forward scattering $\tilde{\mu} = 1$, or $t = 0$. Defining an upper limit of integration $\tilde{\mu}_{\rm max} = \cos \theta_{\rm min} \approx 1 - \theta_{\rm min}^2/2$ in order to regularize the integral and keeping only the leading order terms, we obtain
 \beq\label{eq:Delta1}
    \Delta_1 = 32 \pi^2 \alpha_d^2 \, \ln(4 / \theta_{\rm min}^2) ~ \frac{m_\chi^2 (m_\psi^2 + p^2)}{p^4} \ .
 \eeq

The question is now that of identifying the maximum $\tilde{\mu}_{\rm max}$ or, equivalently, the minimum scattering angle $\theta_{\rm min}$. A prescription commonly used in the literature is to identify this minimum scattering angle with that which results from scattering particles at the largest impact parameter $b_{\rm max}$ possible in the classical limit \cite{McDermott:2010pa,Dvorkin:2013cea,Dvorkin:2019zdi,Buen-Abad:2021mvc}. Since in this limit there is a one-to-one correspondence between the impact parameter $b$ and the scattering angle $\theta$, we can then find $\theta_{\rm min}$ from $b_{\rm max}$. Therefore, we are looking for a length scale beyond which the $\chi\psi \rightarrow \chi\psi$ scattering cannot take place, and we will identify it with the maximum impact parameter $b_{\rm max}$ of the scattering. For thermal plasmas made out of charged particles, such as our DR, there is one such length scale: the Debye screening length $\lambda_{D} = \sqrt{{T_d}/{4 \pi \alpha_d n_\psi}}$ \cite{McDermott:2010pa}. Beyond this distance, the dark charge of the $\chi$ particles is screened by $\psi$ plasma effects, and the scattering effectively turns off. Writing the differential scattering cross section in terms of the impact parameter and also of the scattering amplitude, we obtain
 \beq\label{eq:diff_scat}
    b \frac{d b}{d \tilde{\mu}} = \frac{d \sigma}{d \Omega} =  \frac{1}{4 m_\chi \, \sqrt{p^2 + m_\psi^2} v_\psi}\frac{p}{(2 \pi)^2 4 m_\chi} \overline{\vert \mathcal{M}_{\chi\psi} \vert^2} \ .
 \eeq
Here $v_\psi = p/\sqrt{m_\psi^2 + p^2}$ is the velocity of the $\psi$ particles, which is also the relative velocity of the $\chi$ and $\psi$ particles when the iDM is very non-relativistic.

We can input the expression for the square of the matrix element from \Eq{eq:ampl2} into \Eq{eq:diff_scat}. Note that in the limit$m_\psi \gg p$ we recover the usual formula for Rutherford scattering. Finding the relationship between the impact parameter and the scattering angle is now a matter of solving the differential equation \Eq{eq:diff_scat} for $b(\tilde{\mu})$ with the boundary condition $b(-1) = 0$, which corresponds to a head-on collision resulting in backwards scattering. We find
\bea
    b(\tilde\mu) & = & \frac{1}{2 m_\chi} \sqrt{\frac{A(1+\tilde\mu) + 2 B (1-\tilde\mu) \log( (1-\tilde\mu)/2 )}{1-\tilde\mu}} \ , \\
    \text{with }\quad A & \equiv &  \frac{4 \alpha_d^2 m_\chi^2 \bl( m_\psi^2 + p^2 \br)}{p^4} \ , \\
    \text{and }\quad B & \equiv &  \frac{2 \alpha_d^2 m_\chi^2}{p^2} \ .
\eea
The impact parameter is then maximized for $\tilde\mu_{\rm max} \approx 1 - \theta_{\rm min}^2/2$, which leads to
\beq\label{eq:theta_min}
    \theta_{\rm min} = \frac{2 \alpha_d}{b_{\rm max}} \, \frac{\sqrt{m_\psi^2 + p^2}}{p^2} \ ,
\eeq
In the limit of non-relativistic $\psi$ this reduces to the Rutherford scattering result \cite{McDermott:2010pa}.

We could now substitute \Eq{eq:theta_min} into \Eq{eq:Delta1} and then compute $\Gamma$ from \Eq{eq:Gamma_1}. However, since $\theta_{\rm min}$ in \Eq{eq:theta_min} depends on $p$, the resulting expression would be forbiddingly complicated. Taking advantage of the fact that $\theta_{\rm min}$ appears only inside the logarithm, we instead proceed to compute its {\it thermal average} (which is $p$-independent), and use the result to regulate the logarithm. Denoting this thermal average by $\langle \cdot \rangle$, we have
\beq\label{eq:thermal_theta_min}
    \langle \theta_{\rm min} \rangle = \frac{2 \alpha_d}{b_{\rm max}} \frac{\frac{g_\psi}{2\pi^2} \int \dd p \, p^2 f_{\psi0} ~ \frac{\sqrt{m_\psi^2 + p^2}}{p^2} }{n_\psi} = \frac{2 \alpha_d}{b_{\rm max} T_d} \, \frac{x K_0(x) + K_1(x)}{x K_2(x)} \ ,
\eeq
where we have assumed a Boltzmann distribution for $\psi$. Setting $b_{\rm max} = \lambda_D$ and using the formula for the Debye length, we finally obtain an expression for the argument inside the logarithm,
\beq
 \label{eq:regulator}
    \frac{4}{\langle \theta_{\rm min} \rangle^2} = \frac{\pi}{g_\psi \alpha_d^3} \, \frac{K_2(x)}{2 \bl( x K_0(x) + K_1(x)  \br)^2} \ .
\eeq

Now at last we can compute $\Gamma$. Inserting \Eq{eq:regulator} into \Eq{eq:Delta1} and integrating over $q$ in \Eq{eq:Gamma_1}, we obtain
\beq\label{eq:Gamma_final}
    \Gamma = \frac{4}{3 \pi} \, \alpha_d^2 \, \ln (4/\langle \theta_{\rm min} \rangle^2) \, \frac{T_d^2}{m_\chi} \, e^{-x} \bl[ 2 + x(2+x) \br] \ .
\eeq
Note that in the limit of massless $\psi$ this reduces to $\Gamma = \frac{8}{3\pi} \frac{T_d^2}{m_\chi} \alpha_d^2 \ln(\pi / (g_\psi \alpha_d^3))$. This result differs slightly from the exact massless calculation because we used the Maxwell-Boltzmann distribution. In the exactly massless limit, after accounting for the Fermi-Dirac nature of the $\psi$ particles, we would instead get \cite{Cyr-Racine:2015ihg,Buen-Abad:2017gxg}
\beq
    \Gamma = \frac{2 \pi}{9}\frac{T_d^2}{m_\chi} \alpha^2 \ln(4/\langle \theta_{\rm min} \rangle^2) \ .
\eeq
This result is the same as that obtained in Ref.~\cite{Lesgourgues:2015wza}, except for the argument of the $\log$, which differs due to a different choice of regulators for the forward divergence.

\bibliographystyle{JHEP}
\bibliography{spartacous-bib.bib}

\end{document}